\documentclass[journal=aelccp,manuscript=letter]{achemso}
\usepackage[version=3]{mhchem} 

\usepackage{siunitx}
\graphicspath{{Figures/}}
\usepackage[normalem]{ulem}
\usepackage{amssymb}
\usepackage{pdfpages}
\usepackage{xcolor} 

\newcommand{\sub}[1]{$_{\text{#1}}$}
\newcommand{\kv}[2]{\def\in{#1}\def\v{V}\if\in\v\textit{#1}\sub{#2}\else#1\sub{#2}\fi} 
\newcommand{\kvc}[3]{\def\in{#1}\def\v{V}\if\in\v\textit{#1}\sub{#2}$^{#3}$\else#1\sub{#2}$^{#3}$\fi} 
\makeatletter
\newcommand*{\centerfloat}{
  \parindent \z@
  \leftskip \z@ \@plus 1fil \@minus \textwidth
  \rightskip\leftskip
  \parfillskip \z@skip}
\makeatother

\author{Seán R. Kavanagh}
\affiliation[UCL]{Thomas Young Centre and Department of Chemistry, University College London, 20 Gordon Street, London WC1H 0AJ, UK}
\alsoaffiliation[ICL]{Thomas Young Centre and Department of Materials, Imperial College London, Exhibition Road, London SW7 2AZ, UK}
\author{Aron Walsh}
\affiliation[ICL]{Thomas Young Centre and Department of Materials, Imperial College London, Exhibition Road, London SW7 2AZ, UK}
\alsoaffiliation[Yonsei]{Department of Materials Science and Engineering, Yonsei University, Seoul 03722, Republic of Korea}
\email{a.walsh@imperial.ac.uk}

\author{David O. Scanlon}
\affiliation[UCL]{Thomas Young Centre and Department of Chemistry, University College London, 20 Gordon Street, London WC1H 0AJ, UK}
\alsoaffiliation[Diamond]{Diamond Light Source Ltd., Diamond House, Harwell Science and Innovation Campus, Didcot, Oxfordshire OX11 0DE, UK}
\email{d.scanlon@ucl.ac.uk}

\title[Rapid recombination by cadmium vacancies in CdTe]{Rapid Recombination by Cadmium Vacancies in CdTe}

\abbreviations{Keywords?}

\begin{document}
\begin{abstract} 
CdTe is the largest thin-film photovoltaic technology. Non-radiative electron-hole recombination reduces the solar conversion efficiency from an ideal value of \SI{32}{\%} to a current champion performance of \SI{22}{\%}. The cadmium vacancy (\kv{V}{Cd}) is a prominent acceptor species in \emph{p}-type CdTe; however, debate continues regarding its structural and electronic behavior. Using \emph{ab initio} defect techniques, we calculate a negative-U double-acceptor level for \kv{V}{Cd}, while reproducing the \kvc{V}{Cd}{-1} hole-polaron, reconciling theoretical predictions with experimental observations. We find the cadmium vacancy facilitates rapid charge-carrier recombination, reducing maximum power-conversion efficiency by over \SI{5}{\%} for untreated CdTe --- a consequence of tellurium dimerization, metastable structural arrangements, and anharmonic potential energy surfaces for carrier capture.
\end{abstract}

{\large\textbf{TOC Graphic}}

\begin{center}
    \includegraphics[width=3in]{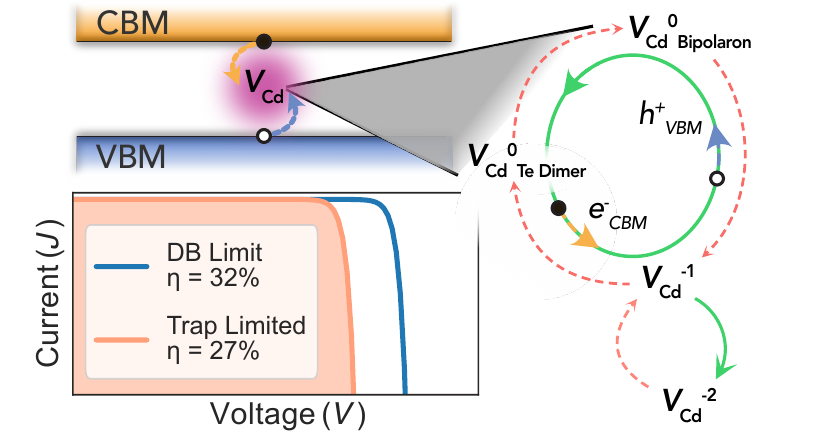}
\end{center}

\medskip
\section{Introduction}
Cadmium telluride (CdTe) is a well-studied thin-film photovoltaic (PV) absorber, being one of few solar technologies to achieve commercial viability.\cite{zidane_optimal_2019}
Its ideal \SI{1.5}{eV} electronic band gap and high absorption coefficient have allowed it to reach record light-to-electricity conversion efficiencies of \SI{22.1}{\%}.\cite{yang_review_2016,noauthor_press_nodate,noauthor_best_nodate}
Given that device architectures and large-scale manufacturing procedures have been highly optimized for this technology --- a result of several decades of intensive research \cite{durose_materials_1999,yang_review_2016} --- further reductions in cost will be heavily-dependent on improvements in photoconversion efficiency.\cite{zidane_optimal_2019,geisthardt_status_2015,yang_review_2016}
Indeed, under the idealized detailed balance model, CdTe has an upper limit of \SI{32}{\%} single-junction PV efficiency (based on its electronic bandgap),\cite{shockley_detailed_1961} indicating that there is still room for improvement.\cite{pan_spin-orbit_2018,geisthardt_status_2015,ma_dependence_2013,reese_intrinsic_2015,burst_cdte_2016}

\medskip

Despite over 70 years of experimental and theoretical research,\cite{castaldini_deep_1998,mathew_photo-induced_2003,emanuelsson_identification_1993,kroger_defect_1977,meyer_anion_1995,meyer_f_1992,lindstrom_high_2015,yang_review_2016,whelan_evidence_1968} the defect chemistry of CdTe is still not well understood. The unambiguous identification of the atomistic origins of many experimentally-observed spectroscopic signatures remains elusive.
Only through clear understanding of defect behavior can strategies be devised to avoid and/or mitigate their deleterious effects on device performance.\cite{huang_perovskite-inspired_2021,li_bandgap_2020,rau_efficiency_2017,green_radiative_2012}

\medskip

At present, market-leading CdTe solar cells employ a Te-rich \emph{p}-type CdTe absorber layer, favoring the formation of Cd vacancies.
Indeed, \emph{undoped} CdTe grown from the melt is typically found to exhibit native \textit{p}-type behavior,\cite{emanuelsson_identification_1993} which has often been attributed to the presence of vacancies in the Cd sub-lattice (and/or Te-on-Cd antisites)\cite{lindstrom_high_2015}.
However, the exact origin of this low intrinsic \textit{p}-type conductivity is still not well understood, with difficulties in definitive measurements\cite{kroger_defect_1977,emanuelsson_identification_1993,meyer_anion_1995,meyer_native_1996}, and discrepancies between models and observations.\cite{yang_review_2016,carvalho_cation-site_2010,menendez-proupin_electronic_2014,menendez-proupin_theoretical_2016,wei_first-principles_2000}
While there is consensus that the cadmium vacancy (\kv{V}{Cd}) is an important acceptor species in CdTe, strong debate has endured regarding its structural and electronic behavior.\cite{takebe_dlts_1982,yang_review_2016,menendez-proupin_electronic_2014,menendez-proupin_theoretical_2016,wei_first-principles_2000,lindstrom_high_2015,shepidchenko_tailoring_2013,reislohner_band-gap_1998,emanuelsson_identification_1993,szeles_trapping_1997}

\section{Theory}
The ability of modern theoretical approaches to accurately describe defect behavior is well-established.\cite{huang_perovskite-inspired_2021,walsh_instilling_2017,park_point_2018,scanlon_acceptor_2009,walsh_instilling_2017}
The use of a sufficiently accurate Hamiltonian is essential for reliable predictions.
For CdTe, using a screened hybrid Density Functional Theory (DFT) functional with spin-orbit coupling (HSE+SOC), we find that the room-temperature experimental bandgap of \SI{1.5}{eV} is reproduced at a Hartree-Fock exchange fraction $\alpha_{\textrm{exx}}=\SI{34.5}{\%}$, a value which also reproduces the experimental lattice constant to within \SI{1}{\%} (see Supporting Information).
For consistency, this model was employed in all structural optimizations and electronic calculations.
\medskip

\section{Results \& Discussion}
\begin{figure}[htb]
    \includegraphics[width=0.68\linewidth]{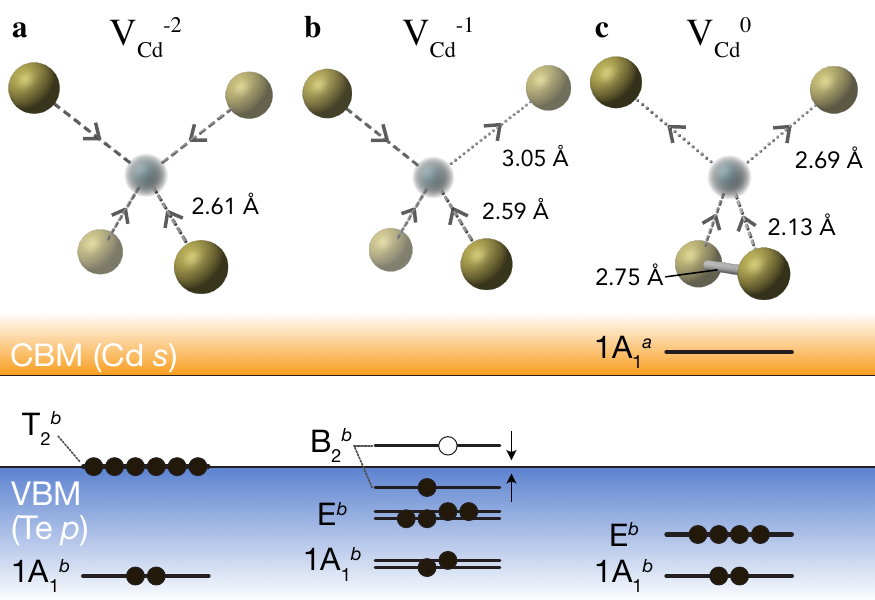}
    \caption{Ground-state structures (\textbf{top}) of the cadmium vacancy in the double-negative (\kvc{V}{Cd}{-2}, \textbf{a}), single-negative (\kvc{V}{Cd}{-1}, \textbf{b}) and neutral (\kvc{V}{Cd}{0}, \textbf{c}) charge states. Tellurium atoms in gold, cadmium vacancy center-of-mass in ocean blue, with each unique Te - \kv{V}{Cd} distance labeled. Also shown are the corresponding electron energy level diagrams (\textbf{bottom}) at the $\Gamma$ point, with character symmetry labels. Superscripts \textit{b} (\textit{a}) refer to (anti-)bonding type interactions.} 
    \label{fig:V_Cd_structures}
\end{figure}

\emph{Cadmium Vacancy: Equilibrium Structures.}
The first step in any theoretical investigation of solid-state defects is the determination of their equilibrium structures.
CdTe crystallizes in the zinc-blende structure (space group $F\bar{4}3m$), thus exhibiting tetrahedral ($T_d$) symmetry at both the Cd and Te sites.
The relaxed geometric configurations upon creation of a cadmium vacancy in the neutral (\kvc{V}{Cd}{0}), single-negative (\kvc{V}{Cd}{-1}) and double-negative (\kvc{V}{Cd}{-2}) charge states are shown in Figure \ref{fig:V_Cd_structures}.
Only the double-negative defect retains the original tetrahedral point-group site symmetry, with a contraction of the neighboring Te atoms from the original bond distance of \SI{2.83}{\angstrom} to \SI{2.61}{\angstrom} from the vacancy center-of-mass. 

The defect site distortions can be rationalized through consideration of the local bonding behavior in a molecular orbital model.\cite{watkins_lattice_1976,watkins_intrinsic_1996}
Removal of a Cd atom (and its two valence electrons) to create a vacancy results in a fully occupied $A_1$ electron level and a $2/3$ occupied $T_2$ level at the Fermi level, arising from the tetrahedral coordination of Te \textit{sp$^3$}-hybrid orbitals. 
In the double-negative case (\kvc{V}{Cd}{-2}), the $T_2$ level becomes fully occupied and thus tetrahedral point symmetry is maintained (Figure \ref{fig:V_Cd_structures}a), with the Te atoms moving closer to the vacancy site to allow for greater hybridization between dangling bonds.

\begin{figure}[t]
\includegraphics[width=\linewidth]{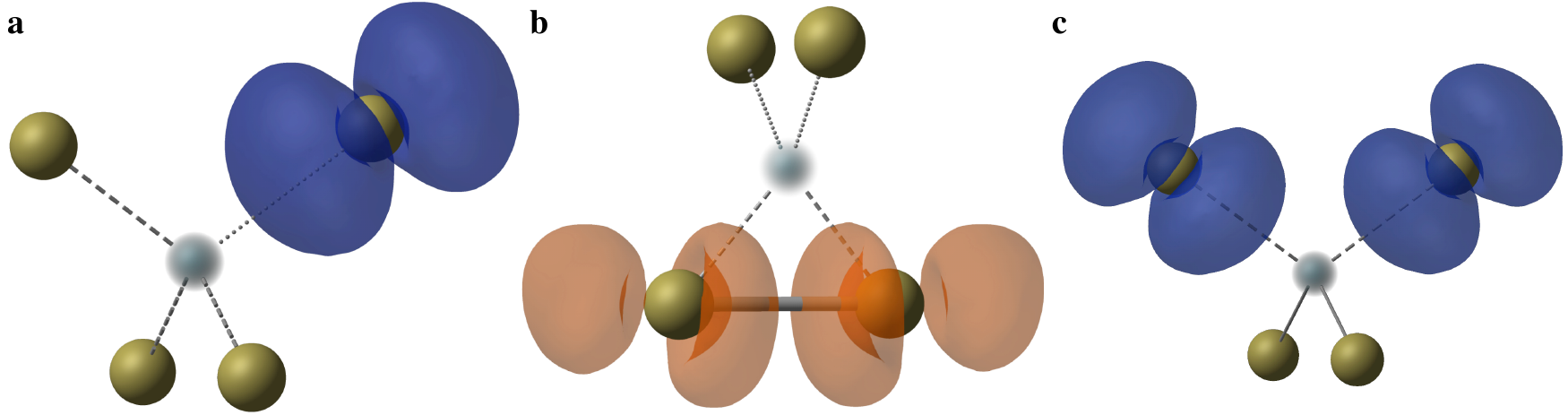}
\caption{Spin-polarized charge-density isosurfaces of the localized hole polaron for the singly-charged defect (\kvc{V}{Cd}{-1}, \textbf{a}), the unoccupied anti-bonding Te dimer state in the neutral vacancy (\kvc{V}{Cd}{0}, \textbf{b}), and the metastable high-spin bipolaron state for the neutral vacancy (\kvc{V}{Cd}{0}$_{\textrm{Bipolaron}}$, \textbf{c}). Tellurium atoms in gold, cadmium vacancy center-of-mass in ocean blue. Isovalues set to \SI{0.006}{e/\angstrom^3} for the polarons (\textbf{a} and \textbf{c}) and \SI{0.002}{e/\angstrom^3} for the dimer state (\textbf{b}).}
\label{fig:Isosurfaces}
\end{figure}

\medskip

For the singly-charged vacancy, the $5/6$ partial occupancy of the $T_2$ level is unstable, undergoing a trigonal Jahn-Teller distortion 
that substantially elongates one of the Te neighbor distances (Figure \ref{fig:V_Cd_structures}b).
In this $C_{3v}$-symmetry vacancy coordination, a positive hole is strongly localized on the Te atom furthest from the vacancy site, as depicted in Figure \ref{fig:Isosurfaces}a, resulting in a paramagnetic defect species.
This $C_{3v}$ polaronic structure of \kvc{V}{Cd}{-1} was experimentally identified in the 1990s, using electron paramagnetic resonance (EPR),\cite{emanuelsson_identification_1993,meyer_anion_1995} but was only reproduced for the first time in a 2015 theoretical study by \citeauthor{shepidchenko_small_2015}\cite{shepidchenko_small_2015}, using the HSE06 functional.
The primary reason why previous \emph{ab initio} works\cite{wei_first-principles_2000,yang_review_2016,du_carrier_2008,chang_symmetrized-basis_2006,lordi_point_2013,biswas_what_2012,carvalho_cation-site_2010} have failed to identify this polaronic ground-state structure for \kvc{V}{Cd}{-1} is the spurious electron self-interaction and consequent over-delocalization inherent in standard (semi-)local DFT functionals.\cite{alberi_2019_2019,oganov_chapter_2018,huang_perovskite-inspired_2021,freysoldt_first-principles_2014}

\begin{figure}[htb]
    \includegraphics[width=0.68\linewidth]{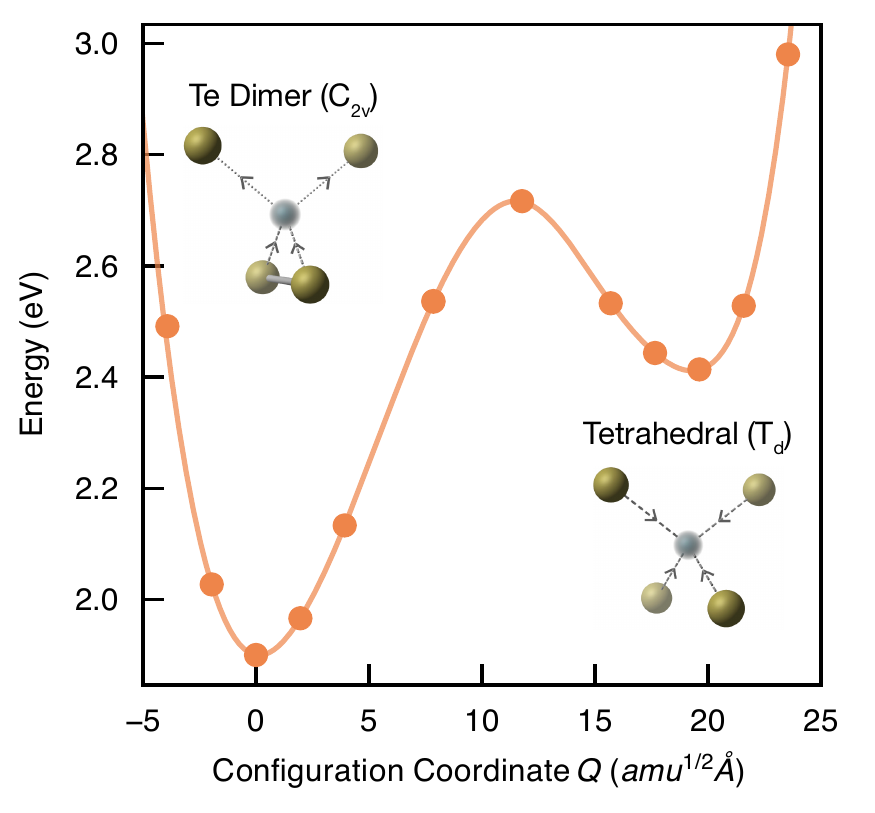}
    \caption{Potential energy surface for \kvc{V}{Cd}{0} along the configurational path from the `Te dimer' ($Q$ = \SI{0}{amu^{1/2}\angstrom}) to tetrahedral ($Q$ $\simeq$ \SI{20}{amu^{1/2}\angstrom}) arrangement. 
    Filled circles represent the calculated formation energies at a given configuration coordinate, and the solid line is a spline fit.
    $Q$ is given in terms of mass-weighted displacement and Te-rich conditions ($\mu_{\text{Te}}=0$) are assumed.}
    \label{fig:Td_to_Dimer_PES_map}
\end{figure}

\medskip 

In the neutral case, we find that the Cd vacancy undergoes strong local relaxation to a $C_{2v}$ structural motif, whereby two Te atoms move significantly closer both to the vacancy site and to each other (\SI{2.75}{\angstrom} separation from an initial \SI{4.63}{\angstrom})(Figure \ref{fig:V_Cd_structures}c).
This yields a Te dimer arrangement with occupied \textit{sp}$^{3}$ $\sigma$-bonding electronic levels deep in the valence band and unoccupied anti-bonding states in the conduction band (Figure \ref{fig:Isosurfaces}b).
Notably, this Te dimerization resembles that observed at low energy surfaces and grain boundaries in CdTe, and has been suggested as a source of fast recombination at these locations.\cite{monch_semiconductor_2001,reese_intrinsic_2015,ahr_flat_2002}
Similar metal-metal dimer reconstructions have been noted for neutral \emph{anion} vacancies in the II-VI semiconductors ZnSe and ZnS,\cite{lany_metal-dimer_2004} occurring here for the \emph{cation} vacancy in CdTe due to the metalloid character of the Te anion.

\begin{figure}[bht]
    \includegraphics[width=0.68\linewidth]{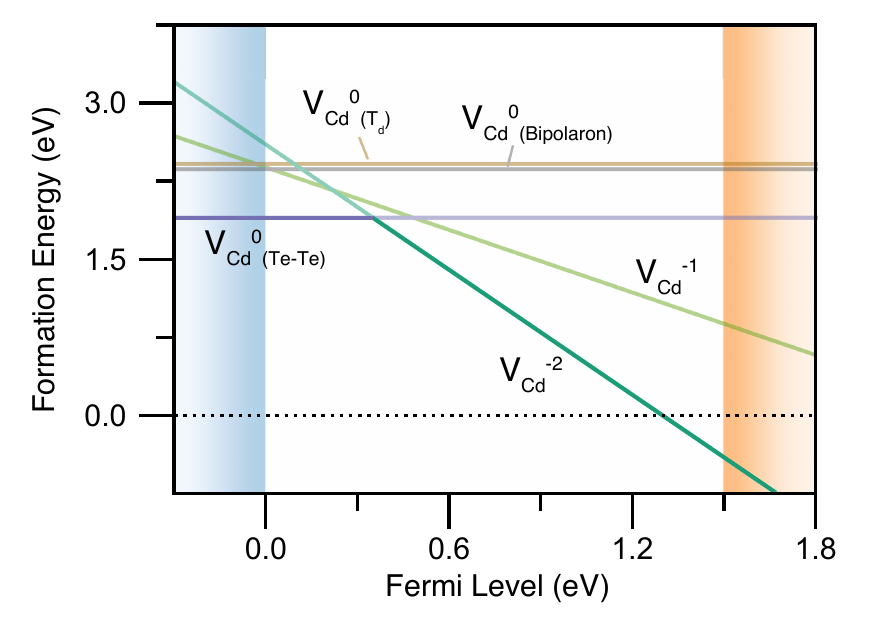}
    \caption{Defect formation energy diagram for the cadmium vacancy in CdTe, under Te-rich conditions ($\mu_{\text{Te}}=0$), with the thermodynamically-favored state for a given Fermi level ($E_F$) shown in saturated color. All locally-stable configurations for the neutral vacancy are included.} 
    \label{fig:V_Cd_LZ_All_Lines}
\end{figure}

This atomic reconstruction reduces the vacancy formation energy by \SI{0.52}{eV}, relative to the tetrahedral solution that has been widely reported\cite{lany_vacancies_2001,chanier_magnetic_2008,wei_first-principles_2000,du_carrier_2008,chang_symmetrized-basis_2006,lordi_point_2013,biswas_what_2012,xu_hybrid_2014} --- Figures \ref{fig:Td_to_Dimer_PES_map} and \ref{fig:V_Cd_LZ_All_Lines}.
As with the $C_{3v}$ Jahn-Teller distortion for \kvc{V}{Cd}{-1}, this Te dimer equilibrium structure of the neutral vacancy has only recently been identified.\cite{lindstrom_high_2015}
The tetrahedral and bipolaron (Figure \ref{fig:Isosurfaces}c) configurations are in fact local minima on the defect potential energy surface (PES), as shown in Figures \ref{fig:Td_to_Dimer_PES_map}, \ref{fig:V_Cd_LZ_All_Lines} and S7. 

The electronic behavior of the Cd vacancy is dramatically affected by Te dimerisation, as the singly-charged state \kvc{V}{Cd}{-1} is consequently predicted to be thermodynamically unfavorable across all Fermi energies (Figure \ref{fig:V_Cd_LZ_All_Lines}).
Accordingly, the vacancy is predicted to act as a so-called negative-U center,\cite{grosse_negative-u_1984,coutinho_characterisation_2020} with a single double-acceptor level at \SI{0.35}{eV} above the valence band maximum (VBM). 
This is in excellent agreement with experimental reports of a single \emph{thermal} ionization level in the bandgap at 0.3–\SI{0.4}{eV} above the VBM (Table S1).\cite{takebe_dlts_1982,szeles_trapping_1997,vul1973investigation,gippius_deep-centre_1974,reislohner_band-gap_1998,scholz_investigations_1999,becerril_effects_2001,kremer_deep_1988}
Moreover, negative-U behavior helps to explain apparent discrepancies between experimental reports of Cd vacancy trap levels, as different techniques can measure either the single-charge ($-2 \rightarrow -1$ and $-1 \rightarrow 0$)
or double-charge transitions ($-2 \rightarrow 0$).\cite{wickramaratne_defect_2018}
The reasons previous theoretical works have not identified this behavior are twofold; namely, incomplete mapping of the defect potential energy surface (overlooking Te-Te dimerization in \kvc{V}{Cd}{0}) and qualitative errors in lower levels of electronic structure theory (destabilizing localized solutions; \textit{viz.} the \kvc{V}{Cd}{-1} small-polaron) --- see Section S6 for further discussion.

\begin{figure}[bht]
    \includegraphics[width=0.68\linewidth]{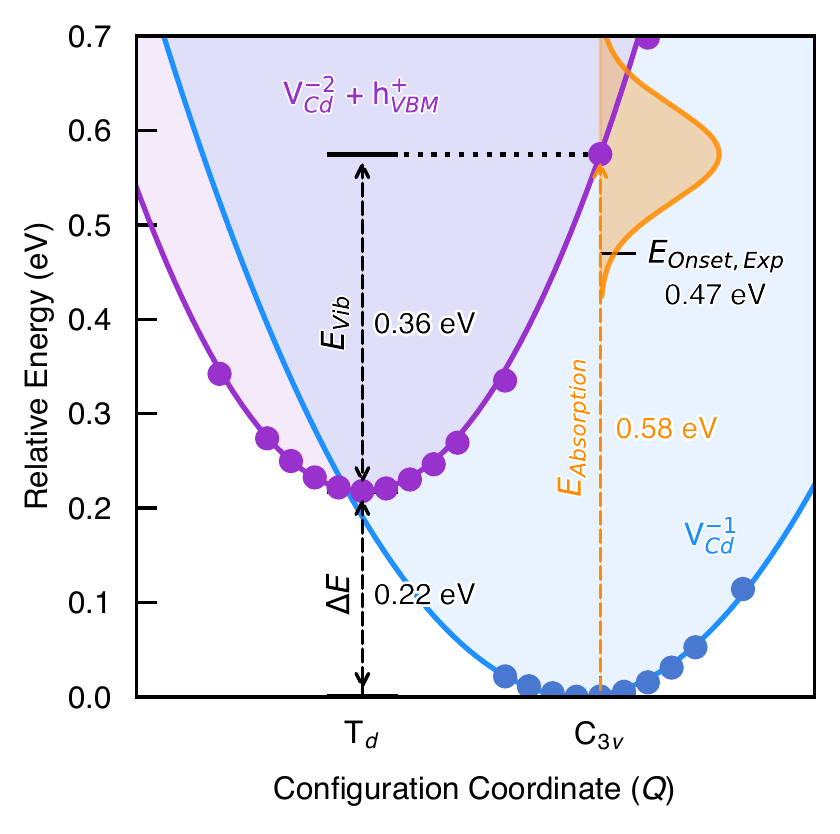}
    \caption{Configuration coordinate diagram for the \kvc{V}{Cd}{-1} $\rightarrow$ \kvc{V}{Cd}{-2} transition, showing the calculated optical excitation ($E_{\,Absorption}$) with vibrational broadening (orange curve), vibrational relaxation ($E_{\,Vib}$), thermodynamic transition ($\Delta E$) and experimental absorption onset ($E_{\,Onset,\, Exp}$) energies.
    The solid lines are harmonic fits to the DFT energies, represented by filled circles.
    \textit{X}-axis labels correspond to the defect point-group symmetry.} 
    \label{fig:V_Cd_-1_CC_diagram}
\end{figure}

\bigskip
\subsection{Optical Response}
\emph{Optical Response.}
The paramagnetic nature of the single negative charge vacancy \kvc{V}{Cd}{-1} (due to the presence of an odd number of electrons)
lends itself to experimental identification through electron spin resonance (ESR/EPR) spectroscopy.
In \citeyear{emanuelsson_identification_1993}, \citeauthor{emanuelsson_identification_1993}\cite{emanuelsson_identification_1993} used photo-ESR to identify the $C_{3v}$ coordination of \kvc{V}{Cd}{-1}, with a localized hole on a Te neighbor as predicted here (Figure \ref{fig:Isosurfaces}a).
After thermal annealing at \SI{750}{\degree C}, they obtained a \textit{p}-type CdTe film with a carrier concentration \textit{p} = \SI{1.2e17}{cm^{-3}}, in excellent agreement with our predicted maximum hole concentration of \textit{p} = \SI{4.2e17}{cm^{-3}} at this temperature (based on calculated intrinsic defect formation energies).
While \kvc{V}{Cd}{-1} is never the lowest energy configuration at equilibrium, we find that Cd vacancies do in fact adopt this charge state under high-temperature p-type growth conditions, as a consequence of energy minimization within the constraint of charge neutrality (to counteract the large hole concentration).

\citeauthor{emanuelsson_identification_1993}\cite{emanuelsson_identification_1993} interpreted a decrease in the \kvc{V}{Cd}{-1} ESR intensity upon irradiation with photons of energy $h\nu > \SI{0.47}{eV}$ as the optical excitation of an electron from the valence band to the ($-/2-$) \kv{V}{Cd} level, to produce \kvc{V}{Cd}{-2} $+\ h_{\,VBM}^+$.
Using the defect structures obtained in our investigations, we calculate the peak energy of this transition as \SI{0.58}{eV}, with vibronic coupling estimated to give a Gaussian lineshape with a FWHM of \SI{0.12}{eV}, yielding good agreement with experiment (Figure \ref{fig:V_Cd_-1_CC_diagram}).

\bigskip
\subsection{Trap-Mediated Recombination}
\emph{Trap-Mediated Recombination.}
To determine the non-radiative recombination activity, electron and hole capture coefficients were calculated for each charge state of the defect.
This approach, building on the developments of \citeauthor{alkauskas_first-principles_2014},\cite{alkauskas_first-principles_2014} uses the \verb|CarrierCapture.jl| package\cite{sunghyun_kim_carriercapturejl_2020} and full details of the calculation procedure are provided in Section S8.
The PES of the defect is mapped along the structural path (configuration coordinate) $Q$ between the equilibrium geometries for a given charge transition, from which nuclear wavefunction overlaps can be determined via the 1D Schrödinger equation.\cite{kim_anharmonic_2019,alkauskas_first-principles_2014}
Electron-phonon coupling is then calculated under static coupling perturbation theory which, in combination with phonon overlaps and scaling factors for charge interaction effects, yields the carrier capture coefficients $C_{p/n}^{\,q}$.

\begin{figure*}[t] 
    \includegraphics[width=\textwidth]{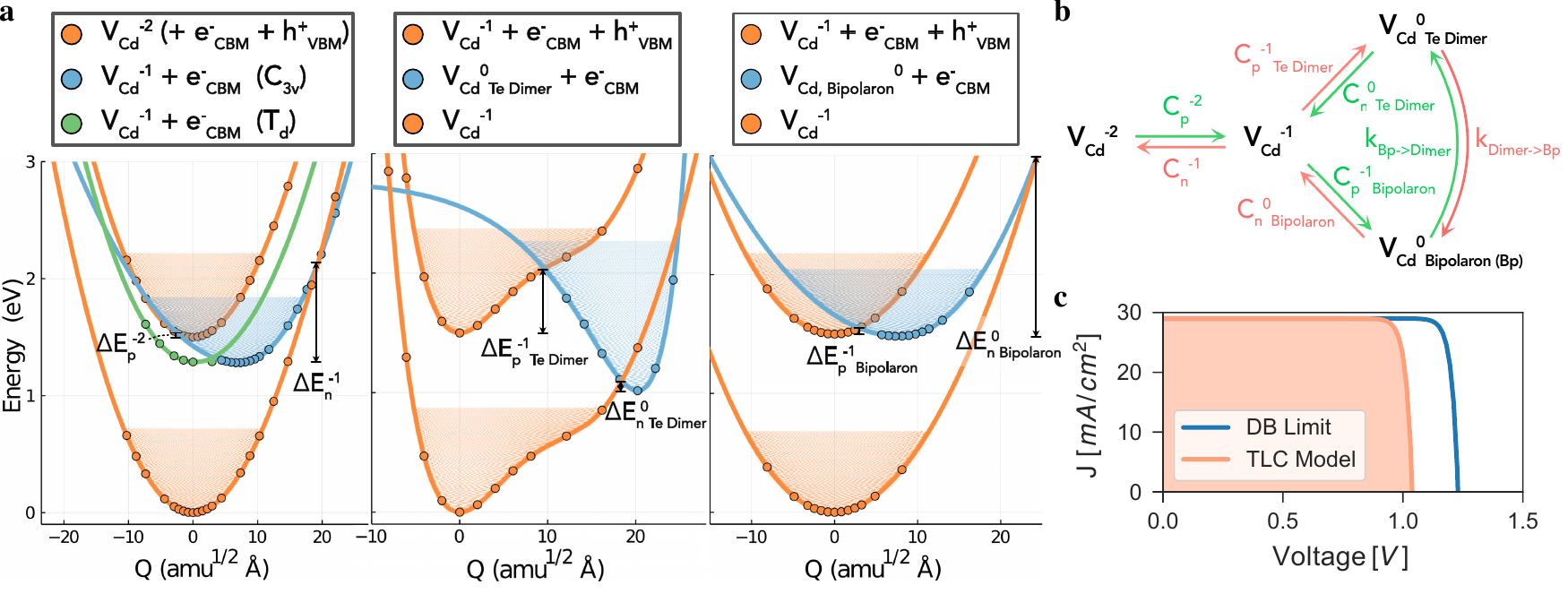}
    \caption{Potential energy surfaces (\textbf{a}) of the $(2-/-)$ (\textbf{left}), $(-/0)_{\textrm{Te\ Dimer}}$ (\textbf{center}) and $(-/0)_{\textrm{Bipolaron}}$ (\textbf{right}) charge transitions for \kv{V}{Cd} in CdTe, with $\Delta E_{p/n}^{\,q}$ denoting the classical energy barrier to hole/electron capture by a vacancy in charge state $q$.
    Filled circles represent calculated energies and the solid lines are best fits to the data.
    The vibrational wave functions are also shown.
    $Q$ is the configurational coordinate path between equilibrium configurations, given in units of mass-weighted displacement.
    (\textbf{b}) Schematic of the  non-radiative recombination mechanism at the cadmium vacancy, with the dominant (rapid) processes colored green.
    (\textbf{c}) $J-V$ curve for an ideal CdTe solar cell, based on the bulk electronic properties and excluding interfacial effects. ``TLC" (Trap-Limited Conversion Efficiency) refers to a device limited by non-radiative recombination at \kv{V}{Cd} (details in text), and ``DB" is the Detailed Balance limit.}
    \label{fig:AllTheCarrierCapture}
\end{figure*}

The energy surfaces for all in-gap \kv{V}{Cd} carrier traps are shown in Figure \ref{fig:AllTheCarrierCapture} and the resulting capture coefficients tabulated in Section S8.
As expected for an acceptor defect with a trap level near the VBM (Figure \ref{fig:V_Cd_LZ_All_Lines}), hole capture is fast while electron capture is slow for the (2-/-) transition, with small and large capture barriers, respectively.
For the \kvc{V}{Cd}{-1} $\rightleftarrows$ \kvc{V}{Cd}{0} transitions, however, the behavior is drastically different to that predicted by a simple quantum defect model.\cite{das_what_2020}
Firstly, hole capture is more rapid than expected, due to the ability of \kvc{V}{Cd}{-1} to transition to the metastable \kvc{V}{Cd}{0}$_{\textrm{Bipolaron}}$ configuration, before relaxing to the \kvc{V}{Cd}{0}$_{\textrm{Te\ Dimer}}$ ground-state.
Secondly, despite the (-/0)$_{\textrm{Te\ Dimer}}$ trap level lying over \SI{1}{eV} below the CBM (Figure \ref{fig:V_Cd_LZ_All_Lines}), typically implying slow electron capture, we in fact find a giant electron capture coefficient.
This unusual behavior is a direct result of the anharmonicity of the PESs at this trap center, accompanied by large electron-phonon coupling, through Te dimer formation. 
These findings provide additional evidence to support Te dimerization at surfaces and grain boundaries in CdTe as a cause of high recombination velocities at these locations.\cite{monch_semiconductor_2001,reese_intrinsic_2015,ahr_flat_2002}
Consequently, the $(-/0)$ \kv{V}{Cd} charge transition is predicted to facilitate rapid electron-hole recombination, proceeding via the \{\kvc{V}{Cd}{-1} $+$ $e_{\,CBM}^-$ $+$ $h_{\,VBM}^+$\} $\rightarrow$ \{\kvc{V}{Cd}{0}$_{\textrm{Bipolaron}}$ $+$ $e_{\,CBM}^-$\} $\rightarrow$ \{\kvc{V}{Cd}{0}$_{\textrm{Te\ Dime}r}$ $+$ $e_{\,CBM}^-$\} $\rightarrow$ \{\kvc{V}{Cd}{-1}\} cycle shown in Figure \ref{fig:AllTheCarrierCapture}b.
Notably, the large capture coefficients for the rapid (green) processes are comparable to the most deleterious extrinsic defects in silicon\cite{macdonald_recombination_2004,peaker_recombination_2012} and the kesterite photovoltaic family.\cite{kim_anharmonic_2019,kim_upper_2020} 
This classifies \kv{V}{Cd} as a `killer center' \cite{stoneham_theory_2001} and demonstrates the potential impediment of this native defect species to the photovoltaic efficiency of untreated CdTe.

To quantify the effect of this recombination channel on CdTe solar cell performance, we calculate the `trap-limited conversion efficiency' (TLC),\cite{kim_upper_2020} which incorporates the effects of defect-mediated non-radiative recombination via the Shockley-Read-Hall model.\cite{shockley_statistics_1952} 
This allows us to set an upper limit on the achievable photovoltaic efficiency in the presence of defects.
As depicted in the current-voltage curve in Figure \ref{fig:AllTheCarrierCapture}c, we find that cadmium vacancies can significantly reduce the open-circuit voltage ($V_{\textrm{OC,\ TLC}} = \SI{1.04}{V}$), minority carrier lifetime ($\tau_e = \SI{29}{ns}$) and thus maximum achievable photovoltaic efficiency from the ideal \SI{32.1}{\%} to \SI{26.7}{\%} (for intrinsic \textit{p}-type CdTe processed under typical anneal temperatures of \SI{600}{\celsius} in a Te-rich atmosphere; see Section S8).
Due to the large hole concentrations in the \textit{p}-type compound, \kvc{V}{Cd}{0}$_{\textrm{Te\ Dimer}}$ will be the dominant state under steady-state illumination, with electron capture by this defect species representing the rate-limiting step:
$$R_{\textrm{Total}}\ \simeq\ R_n^{\,0}\,_{\textrm{Te\ Dimer}}\ =\ n\,C_n^{\,0}\,[\textrm{\kvc{V}{Cd}{0}}_{\textrm{Te\ Dimer}}]$$
Our prediction is a testament to the importance of Cl treatment, strategic impurity doping and Cd-rich growth environments in the fabrication of high efficiency CdTe devices,\cite{metzger_exceeding_2019,yang_enhanced_2015,szeles_trapping_1997,amarasinghe_obtaining_2018,moseley_luminescence_2018,major_low-cost_2014,metzger_time-resolved_2003,park_point_2018,moutinho_effects_1998,ma_dependence_2013,kranz_doping_2013,gessert_research_2013,burst_cdte_2016,komin_effect_2003,kanevce_roles_2017} which contribute to the passivation and reduction of cadmium vacancy populations.
Notably, the recent achievement of open-circuit voltages surpassing the \SI{1}{V} threshold for CdTe solar cells by \citeauthor{burst_cdte_2016}\cite{burst_cdte_2016}, required a switch to an unorthodox strategy of Cd-rich growth conditions and group V anion doping, reducing the formation of \kv{V}{Cd} (and \kv{Te}{Cd}).

\medskip
\section{Conclusions}\label{sec:Conclusions}
In conclusion, we reconcile several longstanding discrepancies between theoretical predictions and experimental measurements for CdTe, predicting both a single double-acceptor level and the $C_{3v}$ \kvc{V}{Cd}{-1} hole-polaron state for the cadmium vacancy in CdTe.
An equilibrium population of cadmium vacancies can facilitate rapid recombination of electrons and holes, reducing the maximum achievable power-conversion efficiency under idealized conditions by over \SI{5}{\%}, for untreated CdTe.
These recombination kinetics primarily arise from both metastable vacancy structures and the Te dimer configuration of \kvc{V}{Cd}{0} which, in addition to producing negative-U behavior, leads to anharmonic carrier capture PESs.
Importantly, these results demonstrate the necessity to include the effects of both metastability and anharmonicity for the accurate calculation of charge-carrier recombination rates in photovoltaic materials.

\begin{acknowledgement}
We thank Dr.~Anna Lindström for valuable discussions regarding polaronic structures for cadmium vacancies, and Dr.~Sungyhun Kim for assistance with \verb|CarrierCapture.jl| calculations.
Seán R. Kavanagh acknowledges the EPSRC Centre for Doctoral Training in the Advanced Characterisation of Materials (CDT-ACM)(EP/S023259/1) for funding a PhD studentship.
We acknowledge the use of the UCL Grace High Performance Computing Facility (Grace@UCL), the Imperial College Research Computing Service, and associated support services, in the completion of this work. 
Via membership of the UK's HEC Materials Chemistry Consortium, which is funded by the EPSRC (EP/L000202, EP/R029431, EP/T022213), this work used the ARCHER UK National Supercomputing Service (www.archer.ac.uk) and the UK Materials and Molecular Modelling (MMM) Hub (Thomas --- EP/P020194 \& Young --- EP/T022213).
\end{acknowledgement}

\begin{suppinfo}
Computational Methods; Bulk Electronic Structure; Vacancy Bonding, Structural \& Electronic Analysis; Discrepancies in Theoretical Studies; Carrier Capture Model, Results \& Analysis, Experimental Identification of Tellurium Dimerization; Defect Electronic Densities of States.
Data produced during this work is freely available at: \url{10.5281/zenodo.4541602}.
\end{suppinfo}

\bibliography{zotero}

\pagebreak


\section*{Supplementary material}

# Supporting Information:

# Rapid Recombination by Cadmium Vacancies in CdTe

Seán R. Kavanagh,[†,‡] Aron Walsh,[*,‡,¶] and David O. Scanlon[*,†,§]

†Thomas Young Centre and Department of Chemistry, University College London, 20 Gordon Street, London WC1H 0AJ, UK

‡Thomas Young Centre and Department of Materials, Imperial College London, Exhibition Road, London SW7 2AZ, UK

¶Department of Materials Science and Engineering, Yonsei University, Seoul 03722, Republic of Korea

§Diamond Light Source Ltd., Diamond House, Harwell Science and Innovation Campus, Didcot, Oxfordshire OX11 0DE, UK

E-mail: a.walsh@imperial.ac.uk; d.scanlon@ucl.ac.uk

## S1    Computational Methods

All calculations were performed using Density Functional Theory (DFT) within periodic boundary conditions through the Vienna *Ab Initio* Simulation Package (VASP).[S1–S3] The screened hybrid DFT exchange-correlation functional of Heyd, Scuseria and Ernzerhof (HSE)[S4] was used for geometry optimizations, calculations of optical dielectric constants, total energies and electronic band structures.[S5] HSE is a range-separated, screened hybrid-DFT functional which incorporates a portion of exact Hartree–Fock exchange $\alpha_{exx}$ for short-range interactions, using a screening parameter of $\omega = 0.11\,\mathrm{bohr}^{-1}$, with the remaining exchange-correlation effects treated by the Generalized Gradient Approximation (GGA) DFT func-

tional PBE.[S6] With the HSE functional, the DFT exchange-correlation energy $E_{xc}$ is given as:

$$
\begin{aligned}
E_{xc}^{HSE}(\alpha_{exx}, \omega) = \quad & \alpha_{exx} E_x^{HF}(\omega) + (1 - \alpha_{exx}) E_x^{PBE,SR}(\omega) \\
& + E_x^{PBE,LR}(\omega) + E_c^{PBE}
\end{aligned}
\tag{1}
$$

where "SR" and "LR" refer to the short and long-range potential components, respectively. To fully account for relativistic effects, spin–orbit interactions were included (HSE+SOC) in all total energy, electronic and optical calculations. Using the projector-augmented wave method, scalar-relativistic pseudopotentials were employed to describe the interaction between core and valence electrons.[S7]

The ionic dielectric response was calculated under Density Functional Perturbation Theory (DFPT) using the PBEsol GGA DFT functional - which has been shown to yield accurate results for this form of calculation,[S8,S9] while the optical response was calculated using the method of Furthmüller *et al.* to obtain the high-frequency real and imaginary dielectric functions.[S8]

A convergence criterion of $0.01\,\mathrm{eV/\mathring{A}}$ was imposed on the forces on each atom during structural optimization, both for the bulk material and defect supercells. Bulk electronic structure calculations were carried out with a two-atom primitive unit cell, using a $12 \times 12 \times 12$ Γ-centered Monkhorst-Pack **k**-point mesh (equivalent to a **k**-point density of $0.1387\,\mathring{A}^{-1}$ in reciprocal space) and a well-converged $450\,\mathrm{eV}$ plane-wave energy cutoff. Charge carrier effective masses were obtained from non-parabolic fitting of the electronic band edges using the *effmass* package,[S10] and electronic band structure diagrams were generated using the *sumo* package.[S11]

For defect calculations, a 64-atom supercell was used, produced from a $2 \times 2 \times 2$ cubic expansion of the conventional zinc-blende CdTe unit cell. The same plane-wave energy cutoff ($450\,\mathrm{eV}$) was employed, with a $2 \times 2 \times 2$ Γ-centered Monkhorst-Pack **k**-point mesh (equivalent to a **k**-point density of $0.24\,\mathring{A}^{-1}$ in reciprocal space). Each defective supercell was relaxed to the same ionic force convergence criteria ($0.01\,\mathrm{eV/\mathring{A}}$) as for bulk structure optimization, with spin-polarization allowed, prior to a static total-energy calculation with the inclusion

of spin-orbit coupling effects. To account for spurious finite-size supercell effects, the Lany-Zunger image charge and potential alignment correction scheme was implemented.[S12] If necessary, Moss-Burstein type band filling corrections were also applied.[S13] Convergence of defect formation energy with respect to supercell size was tested by recalculating the formation energy of the neutral $V_{Cd}{}^0$ and doubly-charged cadmium vacancy $V_{Cd}{}^{-2}$ using a 216-atom supercell ($3 \times 3 \times 3$ cubic expansion of the conventional unit cell), for which the formation energies changed by less than $35\,\mathrm{meV}$ in both cases.

For the calculation of optical transition energies, charge corrections were performed using the GKFO method for vertical defect transitions,[S14] again with a Lany-Zunger-type 2/3 scaling of the point-charge correction energy. Vibronic coupling to yield absorption lineshapes were calculated using the formalism outlined in Ref. S15. Defect concentrations were calculated with the `SC-FERMI` code,[S16] using the calculated formation energies of all intrinsic defects in CdTe — a more detailed discussion of which will be given elsewhere, and Crystal Orbital Hamiltonian Populations (COHP) were calculated using the LOBSTER package.[S17] Anharmonic carrier capture coefficients were calculated using the `CarrierCapture.jl` code,[S18] with electron-phonon coupling matrix elements determined using the method outline in Ref. S19 — further details provided in Section S8.

## S2 Bandgap Corrected Hybrid DFT Functional

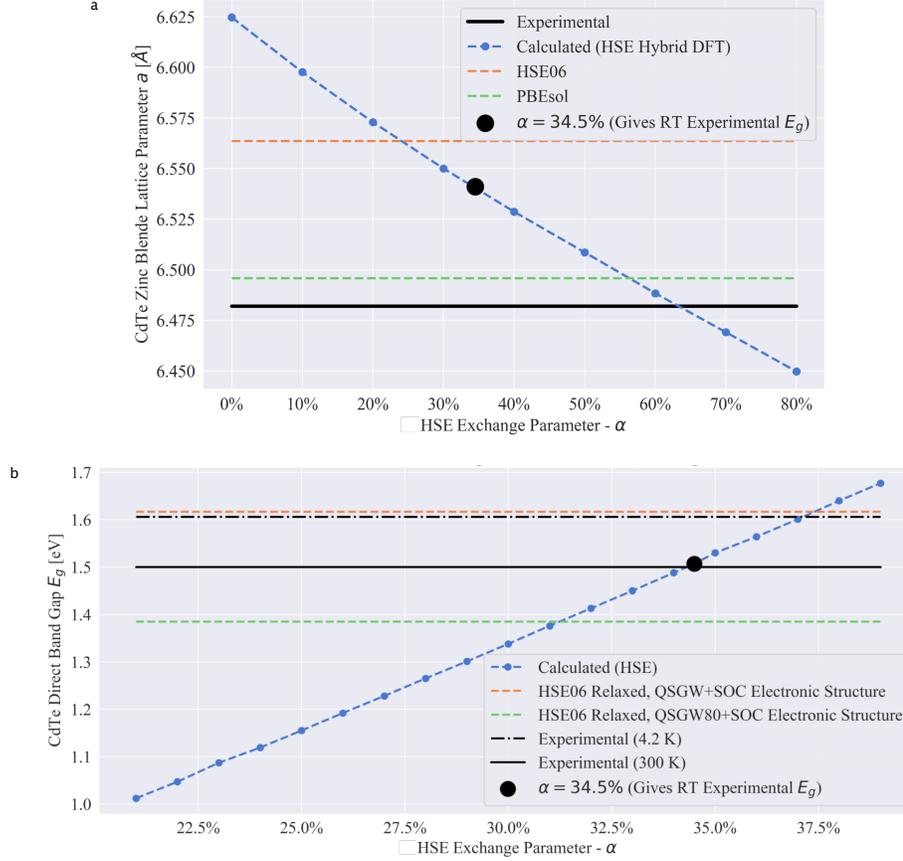

Figure S1: Variation of the calculated cubic lattice parameter (**a**) and fundamental electronic gap (**b**) with respect to HSE exchange fraction $\alpha$ for CdTe, from 0 to 80 % for the lattice constant, and from 20 to 40 % for the bandgap.

To comprehensively and self-consistently investigate the influence of the fraction of exact exchange $\alpha$, employed in the electronic structure model, on the predicted structural and electronic properties of CdTe, bulk structural optimization followed by calculation of the electronic band structure was carried out using the HSE functional with spin-orbit coupling effects (HSE+SOC) for values of $\alpha$ ranging from 0 to 80 %, with the results shown in Figure S1. Quasi-Particle Self-Consistent $GW$ calculations of the electronic bandgap were also performed, for comparison. As expected, the CdTe lattice parameter decreases monotonically,

S-4

while the calculated bandgap monotonically increases, as the fraction of exact exchange is increased, due to the increased localization of the electronic states (decreasing the calculated energy of the occupied valence band, and increasing the energy of the unoccupied conduction band).[S20,S21] We find the the room-temperature (RT) experimental gap of 1.5 eV is reproduced at $\alpha = 34.5\,\%$, and so this value was chosen for the electronic structure model employed in our investigations. This value corresponds to a relaxed lattice constant of 6.54 Å, in good agreement with the experimental value of 6.48 Å (less than 1 % deviation).[S22] We note that Pan et al.[S23] found a similar fraction of exchange (33 %) to reproduce the 1.5 eV experimental bandgap of CdTe[S23] - the slight discrepancy in $\alpha$ likely due to their use of the PBEsol GGA functional for structural optimization.

The same analysis of exchange-property relationships was also applied to other relevant material properties, such as the calculated electron and hole effective masses, spin-orbit splittings and interband transition energies, as shown in Figures S2 and S3.

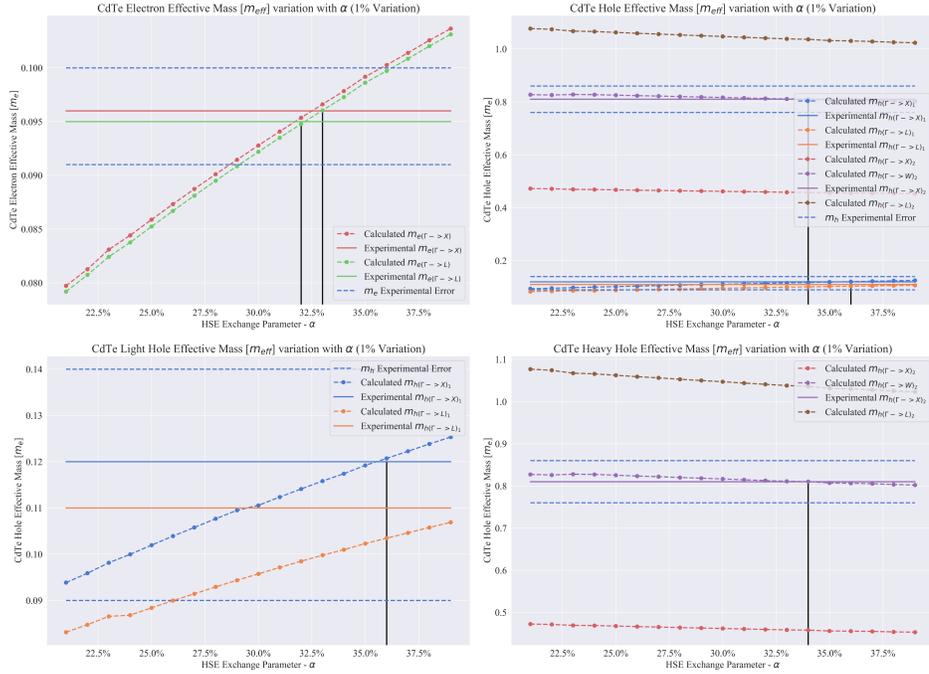

Figure S2: Variation of the calculated charge carrier effective masses for CdTe with respect to HSE exchange fraction $\alpha$, from 20 to 40 %. Experimental values from Madelung,[S24] Strauss[S22] and Thomas.[S25]

S-5

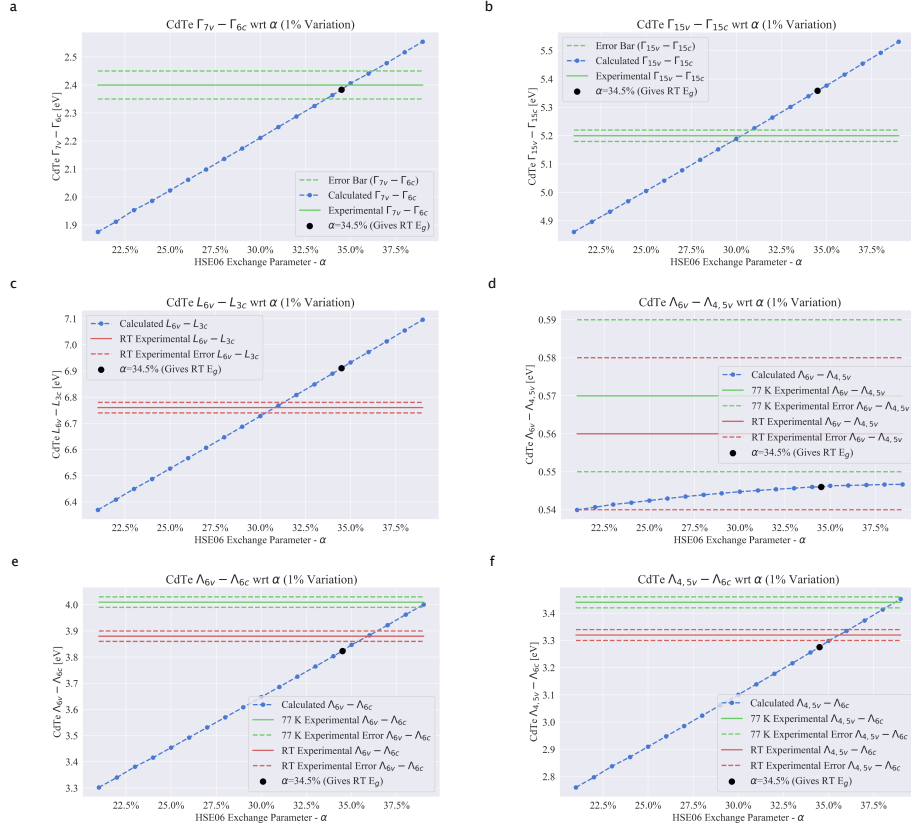

Figure S3: Variation of the calculated spin-orbit splittings and interband transition energies for CdTe with respect to HSE exchange fraction $\alpha$, from 20 to 40%. Experimental values from Madelung,[S24] Strauss[S22] and Thomas.[S25]

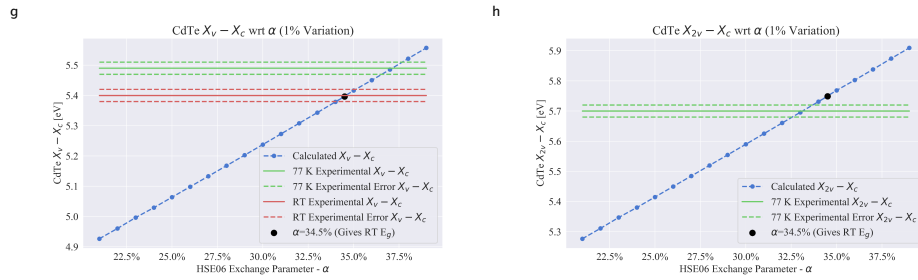

Figure S3: **(contd.)**Variation of the calculated spin-orbit splittings and interband transition energies for CdTe with respect to HSE exchange fraction $\alpha$, from 20 to 40%. Experimental values from Madelung,[S24] Strauss[S22] and Thomas.[S25]

## S3   Bulk Electronic Structure

Using the HSE+SOC functional with 34.5 % exact exchange (HSE(34.5 %) + SOC), the electronic structure of bulk CdTe was calculated, with the results provided in Figure S4.

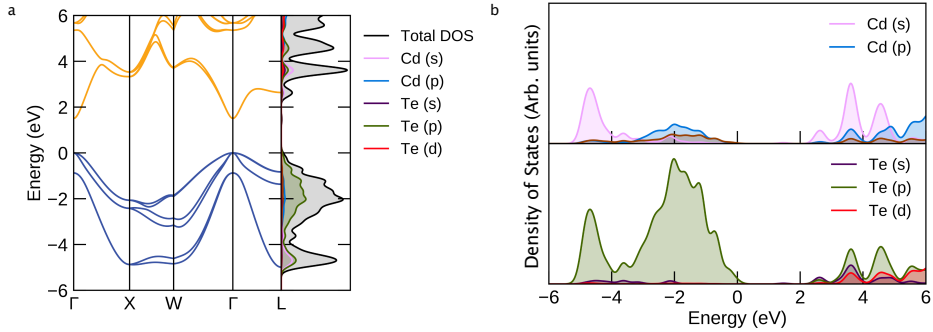

Figure S4: Calculated electronic band structure (**a**) and orbital-projected density of states (**b**) for CdTe, using the HSE(34.5 %)+SOC functional. Valence band in blue, conduction band in orange. VBM set to 0 eV.

CdTe exhibits a direct electronic bandgap, located at $\Gamma = (0, 0, 0)$. Under a standard Molecular Orbital Theory picture, the valence electronic configurations of cadmium and tellurium in their formal oxidation states - $Cd^{2+}$ ($4\,d^{10}5\,s^0$) and $Te^{2-}$ ($5\,s^25\,p^6$) - suggest an valence band maximum comprised of Te $5p$ orbitals and a conduction band minimum formed of Cd $5s$ states. Indeed, analysis of the calculated electronic density of states (Figure S4b) and band orbital characters revealed a conduction band edge dominated by anti-bonding interactions between the Cd $s$ and Te $sp$ hybridized orbitals, while the valence band states primarily arise from the Te $p$ orbitals, as expected.

Notably, the fundamental bandgap was initially calculated as 1.80 eV using the HSE(34.5 %) functional, which decreased to 1.50 eV upon the inclusion of spin-orbit coupling (SOC) effects. This reduction in bandgap is due to the spin-orbit splitting of the Te $p$ - dominated VBM states, as the $5p$ orbitals split into $5p_{1/2}$ and $5p_{3/2}$ states, yielding a VBM upshift of 300 meV. This value is similar to that reported by Pan et al.[S23] (330 meV, with the slight 10 % deviation believed to be a result of their use of the semi-local GGA DFT approximation for structural optimization. The spin-orbit splitting of the VBM states can be witnessed in Figure S4a, where the triply-degenerate valence bands at $\Gamma$ (in the non-SOC case) split into

a pair of degenerate bands at the VBM and a band at $-0.9$ eV below the VBM.

## S4    Neutral Vacancy Bonding Analysis

Crystal Orbital Hamilton Population analysis (COHP) involves partitioning the band-structure energy of a material into a sum of pairwise atomic orbital interactions.[S26,S27] It is defined as:

$$\text{COHP}_{\mu\nu}(E) = H_{\mu\nu}(E)P_{\mu\nu}(E) \tag{2}$$

$$H_{\mu\nu} = \langle \phi_\mu | \hat{H} | \phi_\nu \rangle \tag{3}$$

$$P_{\mu\nu} = \sum_i f_i c_{\mu i}^* c_{\nu i} \delta(\epsilon - \epsilon_i) \tag{4}$$

In essence, COHP analysis indicates bonding, nonbonding, and antibonding energy regions in the electronic density of states, providing a powerful tool for the inspection of chemical bonding behavior in materials. As such, the integrated Crystal Orbital Hamilton Population (ICOHP) for a given pair of atoms or orbitals (i.e. the energy contribution to the band-structure energy) can be viewed as a *measure* of the total energetic contribution of the (anti-)bonding interaction of that pair to the system energy.

$$\text{ICOHP}_{\mu\nu}(E) = \int\limits_{-\infty}^{\epsilon_f} \text{COHP}_{\mu\nu}(E) \, dE \tag{5}$$

A negative value for the (I)COHP indicates an (overall) energy lowering orbital interaction (i.e. bonding), while a positive value indicates anti-bonding type interactions. In the case of the neutral Cd vacancy, the -ICOHP of the two 'dimer' Te atoms is calculated as 4.39 eV, contrasted to a value of 0.028 eV for the Te-Te interaction in bulk CdTe, indicating a strong bonding interaction at this site. Remarkably, this value is in fact larger than the -ICOHP of the per-bond Cd-Te interaction in the bulk (4.02 eV/bond), though of course this comes at the expense of bonding interactions with nearby Cd for the 'dimer' Te atoms. For the tetrahedral and bi-polaron $V_{\text{Cd}}{}^0$ structural arrangements, the maximum Te-Te -ICOHP was calculated as 0.08 eV and 0.05 eV respectively.

The COHP(E) plot for the dimer Te-Te interaction, provided in Figure S5, shows a primar-

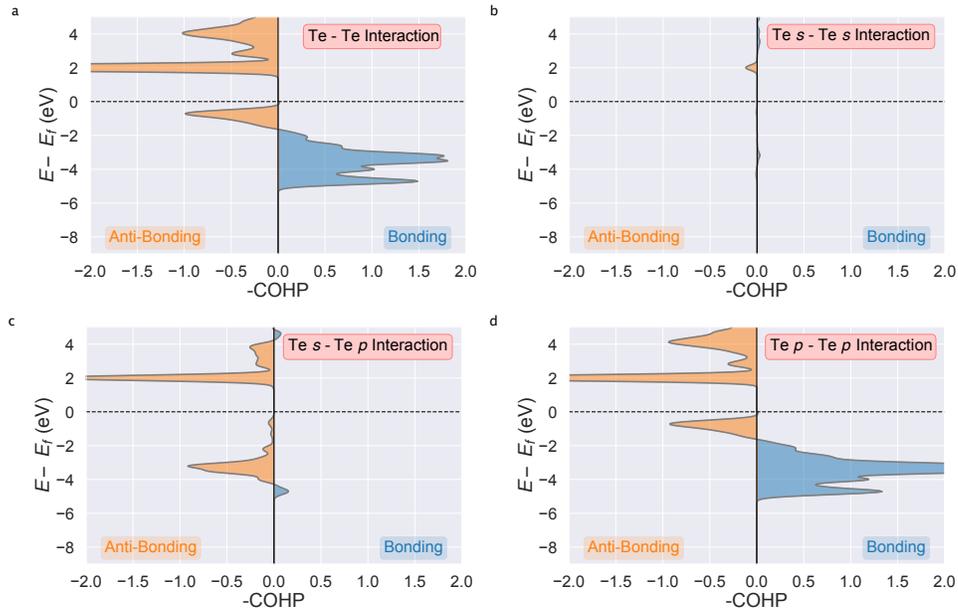

Figure S5: COHP(E) analysis of the Te-Te dimer interaction in $V_{Cd}{}^0$, showing the total (**a**) and $l$-decomposed (**b-d**) contributions.

ily bonding interaction within the VBM, below the Fermi level, and a strong anti-bonding interaction just above the CBM (at $\geq 2\,\mathrm{eV}$).

The electronic density of states for a CdTe supercell containing this defect species is provided in Figure S6, showing a Te $sp$ peak above the CBM, corresponding to the $\sigma$ antibonding state of the Te dimer.

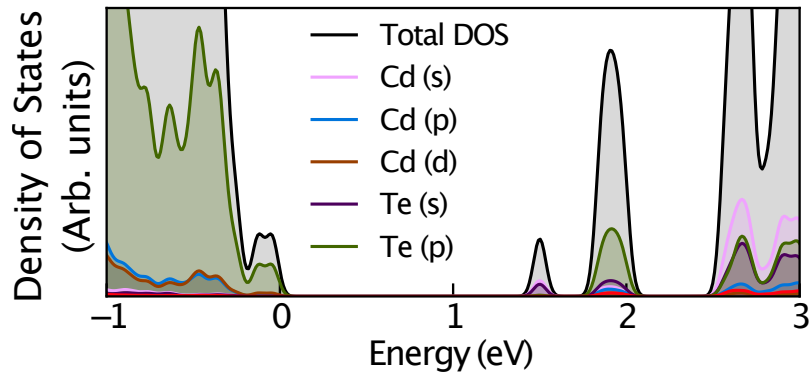

Figure S6: Electronic density of states for a CdTe supercell containing the Te-Te dimer $V_{Cd}{}^0$ species. Fermi level set to 0 eV.

It is worth noting that this Te-dimer structure resembles that observed at low energy surfaces and grain boundaries of CdTe.[S28–S30] A similar metal-metal dimer reconstruction was also found for neutral *anion* vacancies in ZnSe and ZnS.[S31] Here we observe a similar metal-metal dimer for the *cation* vacancy in CdTe, facilitated by the metalloid character of the Te anion.

### S4.1  Bipolaron to Tetrahedral Cd Vacancy PES

The potential energy surface along the configurational coordinate path from the metastable high-spin ($S = 1$) $C_{2v}$ bipolaron solution for $V_{\mathrm{Cd}}{}^0$ to tetrahedral ($T_d$) geometry is shown in Figure S7. Note that the low-spin ($S = 0$) $C_{2v}$ bipolaron structure occurs along this distortion path, producing a divot in the PES at $Q \simeq 6\,\mathrm{amu}^{1/2}\text{Å}$.

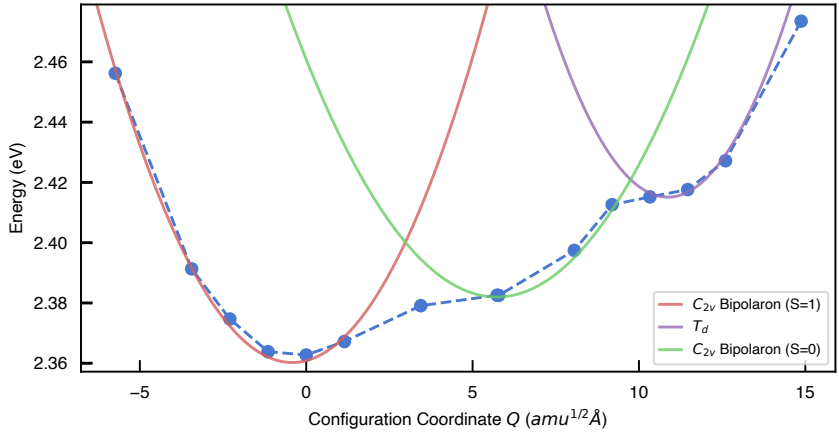

Figure S7: Potential energy surface for $V_{\mathrm{Cd}}{}^0$ along the configurational path from the high-spin $C_{2v}$ Bipolaron configuration ($Q = 0\,\mathrm{amu}^{1/2}\text{Å}$) to tetrahedral ($Q \simeq 20\,\mathrm{amu}^{1/2}\text{Å}$) arrangement. Solid blue circles represent the calculated formation energies at a given configuration coordinate and the quadratic curves are harmonic fits. $Q$ is given in terms of mass-weighted displacement and Te-rich conditions ($\mu_{\mathrm{Te}} = 0$) assumed.

As with similar two-hole polaron cation vacancy defects in II-VI compounds,[S32] the high-spin (HS) state is found to be the lower energy configuration ($\Delta E_{LS-HS} = 40\,\mathrm{meV}$). However, unlike BeO and ZnO (for which this structure has been experimentally-verified as the equilibrium cation vacancy configuration),[S32,S33] this bipolaron state is only metastable

S-10

in CdTe, due to a combination of contributing factors. Firstly, the bipolaronic vacancy is found to decrease in stability (relative to a tetrahedral arrangement) as the size of the group-VI anion increases from O to S to Se to Te, due to decreasing localization of the valence $p$ orbitals.[S32,S34] The lower electronegativity difference and larger atomic radius (of Cd relative to Zn and Be) in CdTe, also act to decrease the stability of this species. Combined with the ability of the metalloidic Te anion to form stable metal-metal dimers, these factors lead to the bipolaron vacancy structure lying $\Delta E_{\text{Bipolaron/Te Dimer}} = 0.47 \, \text{eV}$ higher in energy than the Te-dimer arrangement for $V_{\text{Cd}}{}^0$.

## S5 Thermodynamic Cadmium Vacancy Acceptor Level

Table S1: Experimental reports of the thermodynamic charge transition energy level of the cadmium vacancy acceptor in CdTe. Energies given relative to the VBM.

| Measurement Technique / *DFT Functional* | $\varepsilon(0/2-)$ **[eV]** |
|---|---|
| *HSE(34.5%)+SOC (LZ Correction) - This Work* | **0.35** |
| Deep Level Transient Spectroscopy[S35-S38] (DLTS) | 0.29-0.40 |
| Electronic and Optical Hall Measurements[S39,S40] | 0.30 - 0.37 |
| Dark Current Spectroscopy[S41] | 0.44(1) eV |
| *HSE(34.5%)+SOC (FNV Correction)[S42] - This Work* | **0.47** |

## S6 Discrepancies in Theoretical Studies

The reasons for which previous works have not identified negative-U behavior for the cadmium vacancy are twofold; namely incomplete mapping of the defect potential energy surface (overlooking the Te dimer $V_{\text{Cd}}{}^0$ groundstate) and inherent qualitative errors in lower levels of electronic structure theory (destabilizing localized solutions, namely the $V_{\text{Cd}}{}^{-1}$ small-polaron).

### S6.1 Incomplete Potential Energy Surface Mapping

As demonstrated in Figures 3 and S7, there are often multiple locally-stable configurations for a given defect species. Local optimization algorithms are inherently sensitive to the initial atomic configurations, and thus can become trapped at a local minimum on the PES, failing

to obtain the true global minimum. In the case of $V_{Cd}{}^0$, distortion of the ideal vacancy structure prior to geometry optimization, on the basis of chemical intuition, was required to obtain the equilibrium Te dimer coordination. As evidenced by Figure 4, incorrect prediction of the atomic defect structure can drastically affect the expected electronic behavior; in this case shifting from a single negative-U / (0/2-) charge transition level at 0.35 eV above the VBM, to shallow (0/-) and (-/2-) levels with $V_{Cd}{}^0$ unstable for $E_F$ within the bandgap. The challenge of global optimization is by no means a novel issue. Presently, researchers tend to rely on chemical intuition and heuristics to identify probable minimum-energy atomic arrangements for defect structures,[S43–S45] with the formulation of a general procedure for this crucial step in defect investigations remaining a challenge for the field.

Notably, by sampling the PES via random 'rattling' atomic displacements, we identify another different locally-stable $C_{2v}$ coordination for $V_{Cd}{}^0$, involving parallel spin-polarized holes localized on two of the neighboring Te atoms (which move away from the vacancy site, as the other two move closer)(Figure 2). This two-hole bi-polaron structure is reminiscent of that observed for neutral metal vacancies in other II-VI compounds, such as BeO and ZnO.[S32,S33] While energetically favorable relative to the tetrahedral coordination ($\Delta E_{T_d/Bipolaron}$ = 0.05 eV, Figure S7), this bi-polaron arrangement is only metastable in CdTe (a consequence of the small electronegativity difference and large atomic radii), lying 470 meV higher in energy than the Te-dimer vacancy structure (Section S4.1).

## S6.2   Functional Choice

The judicious choice of exchange-correlation energy functional within DFT is crucial to the accuracy of the electronic structure model.[S44,S46–S49] The singly-charged cadmium vacancy provides a striking example of this requirement for an appropriately high level of theory in defect investigations.

As discussed by Lindström et al.,[S44] all earlier theoretical investigations of defects in CdTe did not identify the experimentally-observed $C_{3v}$ trigonally-distorted $V_{Cd}{}^{-1}$ structure.[S43,S50–S55] In this case, failure to predict this structure cannot be attributed to incomplete sampling of the defect PES as (1) we find the single-negative-charge vacancy to spontaneously relax to this configuration, from an initial *undistorted* vacancy coordination, and (2) several of these

studies explicitly trialled $C_{3v}$-distorted structures,[S43,S51,S54,S56] finding them unstable.

The origin of this discrepancy resides in the fraction of exact Hartree-Fock exchange $\alpha_{exx}$ included in the hybrid functional. Using the common value of $\alpha_{exx} = 25\%$, as in the popular HSE06 functional, the $C_{3v}$ polaronic state is calculated to lie only several meV lower in energy than the $T_d$ solution, depending on the choice of other calculation parameters such as $k$-point sampling and inclusion of SOC (this work and Refs. S34,S54,S56). Consequently, the shallow local minimum of the $C_{3v}$ state becomes near-impossible to locate. Increasing the fraction of exact exchange, as performed in this study (Section S2), favors localized electronic states and thus lowers the energy of the polaronic state relative to the delocalized tetrahedral solution ($\Delta E_{C_{3v}/T_d} = 130$ meV, compared to 37 meV for HSE06[S34]) — resulting in the correct defect behavior.[S57,S58] This is a strong validation of the accuracy of bandgap-corrected hybrid functionals in the investigation and prediction of crystal defect properties.

**Spin-Orbit Coupling** Furthermore, while Ref. S44 obtained the same equilibrium structures as in our investigations, their results still indicated a small stability window of 0.06 eV for $V_{Cd}^{-1}$, thus predicting two defect levels in the bandgap. We believe the origin of this discrepancy with our results is the neglect of spin-orbit coupling (SOC) effects in this study. From analysis of the orbital-projected density of states for both the SOC and spin-polarized non-SOC calculations (Section S7), we observe that the spin-orbit splitting of the singly-occupied $B_2$ $V_{Cd}^{-1}$ level ($\Delta_{SO} \simeq 660$ meV) is less than that of the Te $p$ VBM states ($\Delta_{SO} = 880$ meV), resulting in reduced separation of the VBM and the unoccupied $B_2$ $V_{Cd}^{-1}$ level upon explicit inclusion of spin-orbit effects (See Figure S8 and Section S7 for further detail). Consequently, the spin-stabilization of the occupied $B_2$ $V_{Cd}^{-1}$ electronic state, located within the valence band, is overestimated when SOC is neglected. These observations provide a clear illustration of the necessity for the explicit inclusion of spin-orbit interactions within the electronic structure model for accurate prediction of defect behavior in heavy-atom compounds.

# S7  $V_{Cd}{}^{-1}$ Spin-Orbit Effects

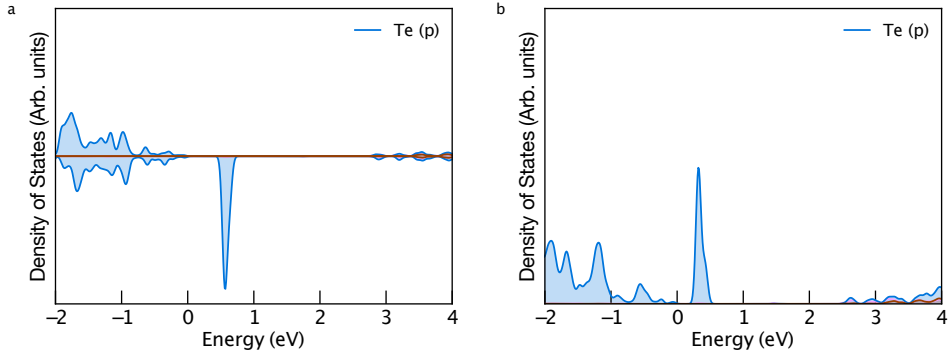

Figure S8: Site-projected electronic density of states for the Te atom furthest from the vacancy site (upon which the hole polaron is localized) in $V_{Cd}{}^{-1}$, with spin-polarized, non-SOC HSE(34.5 %) functional (**a**) and HSE(34.5 %)+SOC (**b**). Fermi level set to 0 eV.

Upon explicit inclusion of spin-orbit coupling (SOC) effects in our electronic structure model (i.e. changing from the spin-polarized, non-SOC HSE(34.5 %) functional to HSE(34.5 %)+SOC), we observe a spin-orbit splitting of $\Delta_{SO} =880$ meV in the Te $p$ VBM states, in good agreement with experimental measurements.[S22,S24,S25] This produces an upshift in energy of the valence band maximum of 300 meV (Sections S2 and S3), as noted by Pan et al..[S23] However, upon inclusion of SOC, the spin-orbit splitting of the singly-occupied $B_2$ $V_{Cd}{}^{-1}$ level is approximately $\Delta_{SO} \simeq 660$ meV, less than that of the Te $p$ VBM states ($\Delta_{SO} =880$ meV) and thus reducing the separation between the VBM and the unoccupied $B_2$ $V_{Cd}{}^{-1}$ level by 200 meV relative to the non-SOC case (Figure S8). The spin-stabilization of the occupied $B_2$ $V_{Cd}{}^{-1}$ electronic state, located within the valence band, is consequently overestimated when SOC is neglected. The reason for the reduced spin-orbit splitting of the $B_2$ $V_{Cd}{}^{-1}$ levels is a lower p-character of this state compared to the VBM states (from orbital-projection analysis of the electron bands). Consequently, we predict the minus charge state to be 130 meV higher in energy at the $\epsilon(0/2-)$ charge transition level than the neutral $V_{Cd}{}^0$ and double-negative $V_{Cd}{}^{-2}$ defects.

S-14

# S8 Carrier Capture

While we predict only a single thermodynamic charge transition level for the cadmium vacancy (Figure 4) — corresponding to the addition or removal of two electrons, this in fact does not correspond to an active charge-carrier capture level. The simultaneous capture of two electrons or holes is highly unlikely, and so it is the sequential capture of single carriers that is relevant to the overall non-radiative recombination behavior. Thus, we identify 4 possible trap levels for the Cd vacancy in CdTe; $(-2/-)$ and $(-/0)_x$ for each of $x =$ Te-Te dimer, Bipolaron and $T_d$ configurations of $V_{Cd}{}^0$. These correspond to intersections of lines in the defect formation energy diagram in Figure 4, where the charge states differ by $+/-1$.

For the theory and implementation of non-radiative carrier capture, we direct the reader to Refs. S59 and S60 respectively. The workflow for each trap level may be crudely summarized as:

- Generate a PES configuration coordinate diagram, by performing a series of singlepoint calculations for atomic structures along the interpolated path between the equilibrium configurations of the two defect charge states.

- Find a best fit to generate the potential energy surfaces and solve the 1D Schrödinger equation, for each PES, to obtain the nuclear (phonon) wavefunctions $\chi_{im}$ and $\chi_{fn}$.

- Calculate the electron-phonon coupling matrix elements $W_{if}$ for the band edge and localized defect single-particle states, under static coupling perturbation theory.

- If applicable, calculate the Sommerfeld factor[S61] $s(T)$ (to account for the Coulombic interaction between the charge carrier and a charged defect) and a scaling factor $f$ to account for charged supercell effects on electron-phonon coupling.

- Calculate carrier capture coefficients $C_{p/n}$ and cross-sections $\sigma_{p/n}$ according to:

$$\tilde{C} = \tilde{V}\frac{2\pi}{\hbar}g|W_{if}|^2\sum_{m,n}w_m|\langle\chi_{im}|\Delta Q|\chi_{fn}\rangle|^2$$
$$\times \delta(\Delta E + m\hbar\Omega_i - n\hbar\Omega_f)$$

(6)

S-15

$$C = s(T)f\tilde{C} \tag{7}$$

$$\sigma = \frac{C}{\langle v_{th} \rangle} \tag{8}$$

where $\tilde{V}$ is the volume of the supercell, $g$ is the degeneracy of the final defect state, $w_m$ is the thermal occupation of vibrational state $m$ at temperature $T$, $\Delta Q$ is the mass-weighted displacement between equilibrium defect configurations, $\Delta E$ is the thermodynamic transition energy between defect charge states, $\Omega_{i/f}$ are the effective vibration energies and $\langle v_{th} \rangle = \sqrt{3k_B T/m^*}$ is the thermal carrier velocity.

The overall rate of electron capture $R_{X,q,n}$ for a defect center $X$ in charge state $q$ is then:

$$R_{X,q,n} = C_{X,q}N_{X,q}n \tag{9}$$

where $N_{X,q}$ is the concentration of defect $X$ in charge state $q$, and $n$ is the electron carrier concentration. An analogous equation holds for the overall rate of hole capture $R_{X,q,p}$.

The $(-/0)_{T_d}$ transition can immediately be ruled out as a potential recombination center, as it presents a transition level $30\,\text{meV}$ below the VBM (Figure 4) and is thus a shallow trap state. Consequently, it will act to capture and emit holes, affecting charge transport but not facilitating recombination.

The calculated PESs (configuration coordinate diagrams) for the $(2-/-)$, $(-/0)_{\text{Te Dimer}}$ and $(-/0)_{\text{Bipolaron}}$ transitions are shown in Figure S9.

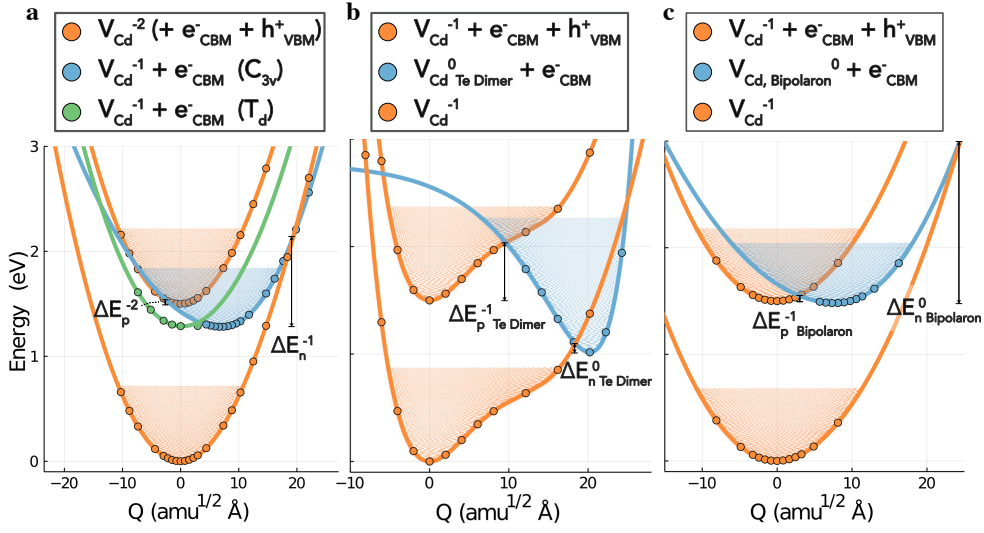

Figure S9: Potential energy surfaces of the (**a**) $(2-/-)$, (**b**) $(-/0)_{\text{Te Dimer}}$ and (**c**) $(-/0)_{\text{Bipolaron}}$ charge transitions for $V_{\text{Cd}}$ in CdTe, with $\Delta E_{p/n}^q$ denoting the classical energy barrier to hole/electron capture by a vacancy in charge state $q$. Filled circles represent calculated energies and the solid lines are best fits to the data. The vibrational wave functions, determined via the 1D Schrödinger equation, are also shown. $Q$ is the configurational coordinate path between equilibrium configurations, given in terms of mass-weighted displacement.

Note that the metastable tetrahedral $T_d$ arrangement of $V_{\text{Cd}}^{-1}$ is also shown in Figure S9a, which essentially corresponds to a $V_{\text{Cd}}^{-2}$-like defect surrounded by a hole in a shallow perturbed-VBM state. The lack of intersection between the PES of this state and that of $V_{\text{Cd}}^{-2}$ results in a tunneling-dominated capture process, with a rate much lower than for the $V_{\text{Cd}}^{-2} + h_{VBM}^+ \rightarrow V_{\text{Cd}}^{-1}(C_{3v})$ transition (calculated difference of $\sim 10$ orders of magnitude), and so it is not discussed further.

Also interesting to note that the PESs for the $(2-/-)$ and $(-/0)_{\text{Bipolaron}}$ are quasi-harmonic, as is typically the case and often assumed in calculations of non-radiative capture rates, while those of the $(-/0)_{\text{Te Dimer}}$ level are not. For the $V_{\text{Cd}}^{-1}$ PES, this is explained by the presence of the nearby metastable $T_d$ configuration (located $\Delta Q = 7.35 \, \text{amu}^{1/2} \text{Å}$ away in configurational space), which merges with the PES of the $C_{3v}$ state to form the adiabatic energy surface shown in Figure S9b. Similar behavior has been witnessed for the so-called $d^0$ $\text{Si}_{\text{Ga}}$ center in GaAs.[S60] For the $V_{\text{Cd}}^{0}_{\text{Te Dimer}}$ PES, we instead find a Morse potential to give a best fit to the data. This is because the structural distortion for $V_{\text{Cd}}^{0}$ here primarily

corresponds to localized bond breaking of the Te-Te dimer, and so the PES resembles an diatomic interaction.

The room temperature Sommerfeld parameters for negatively charged defect centers in CdTe — which depend on the carrier effective masses (determined using `effmass`[S10] in conjunction with the calculated bulk electronic structure (Section S3)) and high-frequency dielectric constant — are provided in Table S2. $s(T) > 1$ indicates an attractive interaction, thus enhancement of the capture rate, while $s(T) < 1$ indicates a repulsive interaction and suppression of the capture rate.

Table S2: Sommerfeld parameters $s(T)$ for negatively charged defect centers in CdTe, at temperature $T = 300\,$K.

| Defect Charge | Carrier (Charge) | Sommerfeld Parameter $s(T = 300K)$ |
|---|---|---|
| -1 | Electron (-1) | 0.0689 |
| -1 | Hole (+1) | 9.71 |
| -2 | Hole (+1) | 19.40 |

Capture cross sections $\sigma_{n/p}$ were calculated according to Equation 8, using the calculated values for carrier effective mass (obtained from non-parabolic fitting of the band edges (Section S3, using the `effmass`[S10] package). The overall capture coefficients and cross-sections at room temperature, as well as electron-phonon coupling matrix elements $W_{if}$, for each $V_{Cd}$ transition level, are provided in Table S3.

Table S3: Electron-phonon coupling $W_{if}$, capture coefficients $C_{n/p}$ and cross-sections $\sigma_{n/p}$ for the $(2-/-)$, $(-/0)_{\text{Te Dimer}}$ and $(-/0)_{\text{Bipolaron}}$ $V_{Cd}$ centers in CdTe, at temperature $T = 300\,$K.

| Transition Level | Carrier (Charge) | $W_{if}$ [eV/amu$^{1/2}$Å] | $C(T = 300K)$ $[cm^3/s]$ | $\sigma(T = 300K)$ $[cm^2]$ |
|---|---|---|---|---|
| $(2-/-)$ | Electron (-1) | $7.9 \times 10^{-4}$ | $2.6 \times 10^{-23}$ | $7.0 \times 10^{-31}$ |
| $(2-/-)$ | Hole (+1) | $3.9 \times 10^{-5}$ | $5.1 \times 10^{-11}$ | $2.4 \times 10^{-18}$ |
| $(-/0)_{\text{Te Dimer}}$ | Electron (-1) | $2.6 \times 10^{-1}$ | $2.6 \times 10^{-6}$ | $7.1 \times 10^{-14}$ |
| $(-/0)_{\text{Te Dimer}}$ | Hole (+1) | $2.1 \times 10^{-2}$ | $3.9 \times 10^{-13}$ | $1.8 \times 10^{-20}$ |
| $(-/0)_{\text{Bipolaron}}$ | Electron (-1) | $3.5 \times 10^{-4}$ | $5.4 \times 10^{-21}$ | $1.5 \times 10^{-28}$ |
| $(-/0)_{\text{Bipolaron}}$ | Hole (+1) | $3.1 \times 10^{-2}$ | $4.4 \times 10^{-6}$ | $2.1 \times 10^{-13}$ |

The variation of the capture coefficients $C_{n/p}$ with temperature $T$, for each transition

level, are shown in Figures S10 and S11.

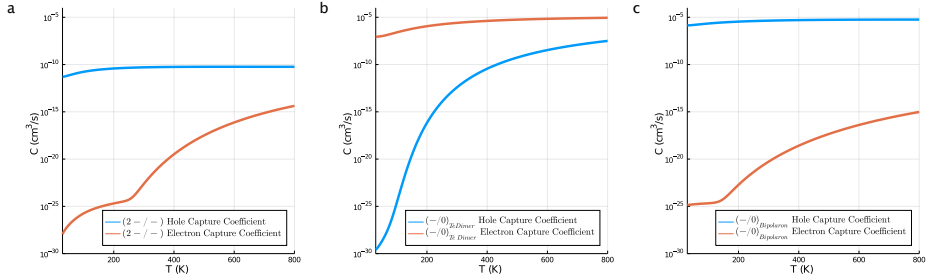

Figure S10: Carrier capture coefficients $C$ as a function of temperature $T$ for the (**a**) $(2-/-)$, (**b**) $(-/0)_{\text{Te Dimer}}$ and (**c**) $(-/0)_{\text{Bipolaron}}$ transition levels of $V_{\text{Cd}}$ in CdTe.

## S8.1 $V_{\text{Cd}}$ — (2-/-)

The hole capture coefficient $C_p$ of the $(2-/-)$ level shows a near-negligible dependence on temperature above $T = 200\,\text{K}$. This is due to the near-zero classical energy barrier $\Delta E = 0.036\,\text{eV}$ for capture (the difference in energy between the PES minimum and the point of intersection with the PES of the other charge state) between the PES of $V_{\text{Cd}}^{-2}$ + $e_{CBM}^-$ + $h_{VBM}^+$ and $V_{\text{Cd}}^{-1}$ + $e_{CBM}^-$ (Figure S9a), rendering it an activationless process. The small activation barrier $\Delta E = 0.036\,\text{eV}$ and near-parallel PESs would suggest an extremely large hole capture coefficient ($C_p^{2-} > 10^{-8}\,\text{cm}^3/\text{s}$) for $V_{\text{Cd}}^{-2}$ (as typically expected for a trap level close to the VBM — Figure 4), however we find the hole capture rate to in fact be limited by relatively weak electron-phonon coupling ($W_{if}$ in Table S3) in this case, yielding a moderate capture rate $C_p^{2-}(T = 300K) = 5.1 \times 10^{-11}\,\text{cm}^3/\text{s}$.

The electron capture coefficient $C_n$, on the other hand, shows a marked dependence on temperature. The reason for this is the domination of tunneling-mediated capture at low temperature, with Arrhenius-like capture behavior only kicking in at $T \sim 250\,\text{K}$, due to the large energy barrier of $0.855\,\text{eV}$ (Figure S9a). It is this large energy barrier, in combination with a relatively small electron-phonon coupling, which yields extremely slow electron capture kinetics at room temperature for $V_{\text{Cd}}^{-1}$ (Table S3).

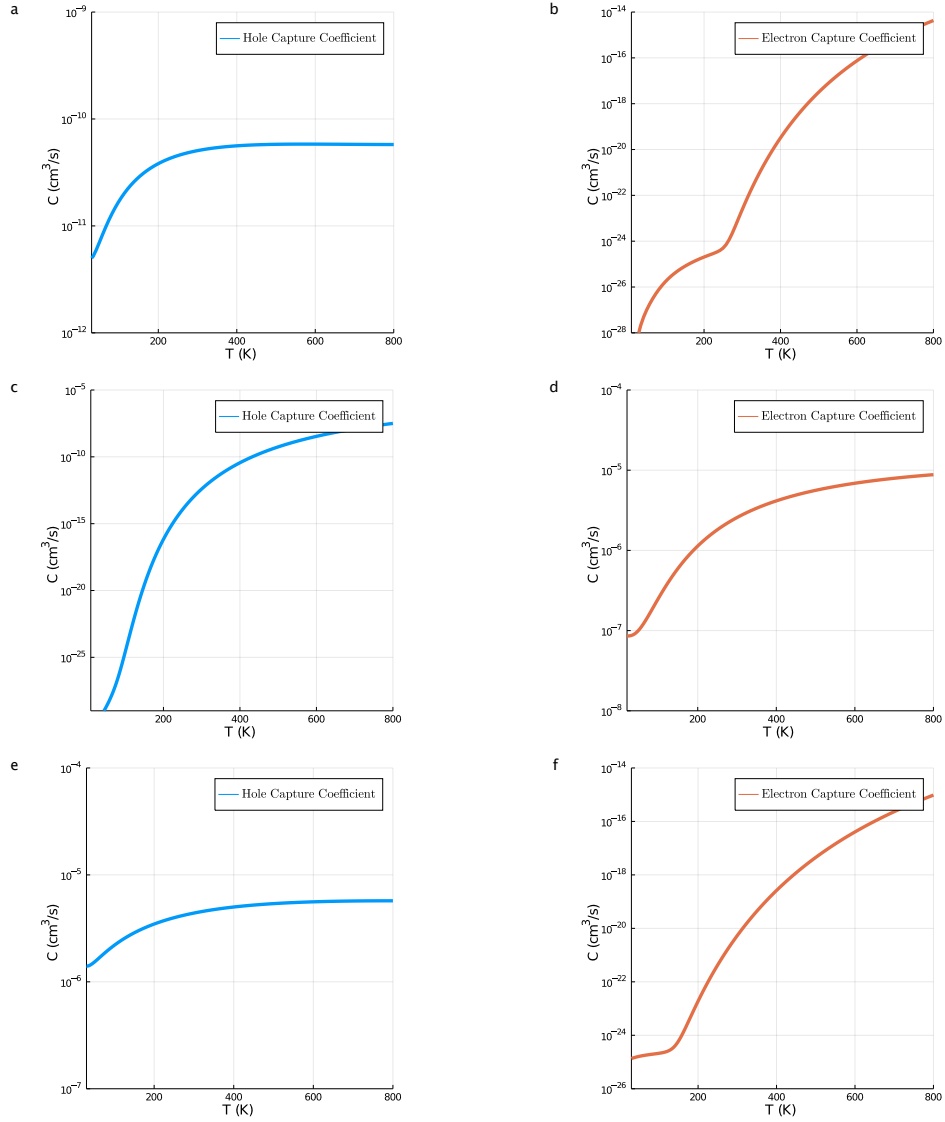

Figure S11: Hole and electron capture coefficients ($C_p$ and $C_n$) as a function of temperature $T$ for the (**a,b**) $(2-/-)$, (**c,d**) $(-/0)_{\text{Te Dimer}}$ and (**e,f**) $(-/0)_{\text{Bipolaron}}$ charge transitions of $V_{\text{Cd}}$ in CdTe. Beware the differing y-axis ranges when considering temperature dependence of the various trap levels.

## S8.2 $V_{\text{Cd}}$ — (-/0)$_{\text{Te Dimer}}$

For the $(-/0)_{\text{Te Dimer}}$ level, the behavior is slightly different, due to the anharmonic PESs (Figure S9b) and different energy barriers. At room temperature, the hole capture rate remains relatively low despite strong electron-phonon coupling, due to the sizable energy barrier $\Delta E = 0.538\,\text{eV}$. Interestingly, the hole capture coefficient $C_p$ shows a strong Arrhenius temperature dependence, with an absence of quantum mechanical tunneling at low temperature. This arises from the large anharmonic lattice relaxation (i.e. Morse-like potential) of the $V_{\text{Cd}}{}^0$ defect (itself a consequence of Te dimer formation), which results in negligible overlap of vibrational wave functions below the classical energy barrier. In other words, Te dimer formation yields a strong separation of the PES minima in configurational space ($\Delta Q = 20.22\,\text{amu}^{1/2}\text{Å}$ compared to 7.35 and $8.12\,\text{amu}^{1/2}\text{Å}$ for $(2-/-)$ and $(-/0)_{\text{Bipolaron}}$), preventing quantum tunneling. Thus, calculated hole capture coefficient decreases exponentially, even at extremely low temperatures $T < 50\,\text{K}$.

In contrast, electron capture occurs with a large overlap of vibrational wave functions, once the small energy barrier ($\Delta E = 0.083\,\text{eV}$) has been overcome, explaining the consistent weak exponential temperature dependence from $T > 50\,\text{K}$ onwards (Figures S10b and S11d). Despite the $(-/0)_{\text{Te Dimer}}$ trap level lying over $1\,\text{eV}$ below the CBM (Figure 4), typically implying slow electron capture, we in fact find an enormous electron capture coefficient. This is a direct result of the anharmonicity of the PESs at this trap center, and is exacerbated by large electron-phonon coupling — arising from the strong local distortion associated with Te dimer formation. Consequently, the harmonic approximation commonly applied for the calculation of carrier capture coefficients, while appropriate for the $(2-/-)$ and $(-/0)_{\text{Bipolaron}}$ centers, grossly fails in this case. This behavior is crucial to the non-radiative recombination activity of the cadmium vacancy as, without it, there would be negligible electron capture at this defect species ($C_n(T = 300K) < 10^{-20}\,\text{cm}^3/\text{s}$ for the $(2-/-)$ and $(-/0)_{\text{Bipolaron}}$ levels) and thus negligible electron-hole recombination.

Moreover, this behavior has important implications for other defects in CdTe, namely tellurium interstitials (Te$_\text{i}$) and tellurium-on-cadmium antisites (Te$_{\text{Cd}}$), which we have found

to exhibit Te dimer structures and high concentrations in $p$-type CdTe. Calculation and analysis of carrier capture kinetics at these trap centers is currently underway.

## S8.3 $V_{\mathbf{Cd}}$ — (−/0)$_{\mathbf{Bipolaron}}$

Finally, for the (−/0)$_{\text{Bipolaron}}$ transition level, the behavior is quite similar to the (2 − /−) level for both the hole and electron capture processes, due to the similarity of the PESs in Figures S9. The only difference is that the hole capture coefficient $C_p$ is about 5 orders of magnitude larger for (−/0)$_{\text{Bipolaron}}$ (Table S3) — due to stronger electron-phonon coupling (Table S3), with almost identical $T$ dependence (Figures S11a,e).

For electron capture, both the magnitude and $T$ dependence of electron capture are essentially identical, due to similar energy barriers (Figure S9) and electron-phonon coupling (Table S3).

## S8.4 $V_{\mathbf{Cd}}{}^{\mathbf{0}}{}_{\mathbf{Bipolaron}} \rightarrow V_{\mathbf{Cd}}{}^{\mathbf{0}}{}_{\mathbf{Te\ Dimer}}$

To estimate the rate of transformation from $V_{\text{Cd}}{}^{0}{}_{\text{Bipolaron}}$ to $V_{\text{Cd}}{}^{0}{}_{\text{Te Dimer}}$, we can invoke Transition State Theory,[S62] which gives a reaction rate of:

$$k \ = \ \nu\, g \exp(-\frac{\Delta E}{k_B T}) \tag{10}$$

where $\nu$ is the attempt frequency, $g$ is the ratio of the degeneracies of the final and initial states and $\Delta E$ is the activation energy barrier. Using the Nudged Elastic Band (NEB) method,[S63] we calculate an upper limit to the barrier of this transition as $\Delta E = 0.30\,\text{eV}$. This value, alongside $\nu = 1.45\,\text{THz}$ (from the interpolated PESs along the linear path between configurations) and $g = 1$ (both configurations have $C_{2v}$ point-group symmetry with 2 sets of 2 equivalent Te atoms), gives a room-temperature transition rate $k_{\text{Bp}\rightarrow\text{Te Dimer}} = 1.32 \times 10^7\,\text{s}^{-1}$. Note that this calculation involves the approximation of equal entropies for the bipolaron and Te dimer states. As expected, this transition occurs more rapidly than both the competing electron capture process:

$$R_n^{0,\,Bipolaron}/[V_{\text{Cd}}{}^{0}{}_{\text{Bipolaron}}] \ = \ C_n^0\, n \ \simeq \ (5.4 \times 10^{-21}\,\text{cm}^3/\text{s})(10^{12}\,\text{cm}^{-3}) \ \simeq \ 10^{-8}\,\text{s}^{-1}$$

and subsequent electron capture process:

$$R_n^{0, Te \ Dimer} / [V_{Cd}{}^0{}_{Te \ Dimer}] \ = \ C_n^0 \, n \ \simeq \ (2.6 \times 10^{-6} \, \mathrm{cm^3/s})(10^{12} \, \mathrm{cm^{-3}}) \ \simeq \ 10^6 \, \mathrm{s^{-1}}$$

Thus it is reasonable to assume that, upon hole capture by $V_{Cd}{}^{-1}$ to form $V_{Cd}{}^0{}_{Bipolaron}$, the metastable neutral vacancy will transform to $V_{Cd}{}^0{}_{Te \ Dimer}$ before electron capture can take place, and that the subsequent electron capture process will be the rate-determining step in the $(-/0)$ $V_{Cd}$ recombination cycle.

## S8.5 Total $V_{Cd}$ Recombination Kinetics

The presence of multiple charge transition levels in the gap means the overall capture kinetics of the $V_{Cd}$ center is governed by a set of coupled rate equations describing the individual capture processes:

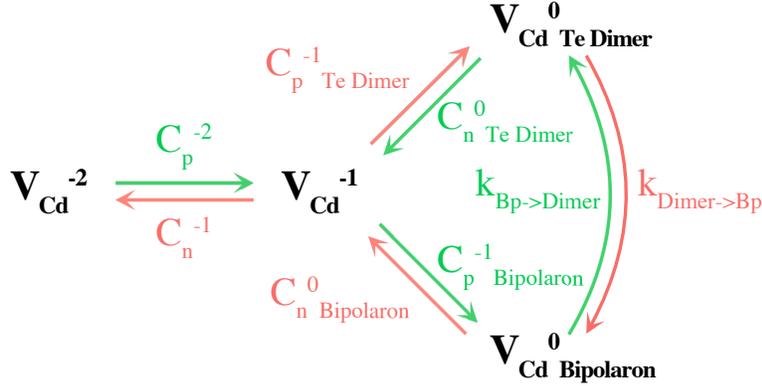

Figure S12: Schematic of the overall non-radiative recombination mechanism at the cadmium vacancy center, where $k_{Bp->Dimer}$ is the transition rate from the bipolaron to Te dimer configuration for $V_{Cd}{}^0$ and the dominant, rapid processes are colored green.

Notably, the large capture coefficients for the rapid (green) processes are comparable to the most deleterious extrinsic defects in silicon[S64,S65] and the kesterite (CZTS) family of thin-film photovoltaics,[S60,S66] classifying them as 'killer' defects[S67] demonstrating the potential impediment of this native defect species to the device efficiency of untreated CdTe.

Under steady-state conditions, each microscopic capture reaction is in quasi-equilibrium:

$$R_p^q \left(= C_p^q \left[D^q\right] p - e_p^{q+1} \left[D^{q+1}\right]\right) \;=\; R_n^{q+1} \left(= C_n^{q+1} \left[D^{q+1}\right] n - e_n^q \left[D^q\right]\right) \tag{11}$$

where $R_p^q$ is the net hole trapping rate of defect $D$ in charge state $q$, $p$ and $n$ are the hole and electron carrier concentrations, $e_p^{q+1}$ and $e_n^q$ are the hole and electron emission coefficients, and $\left[D^q\right]$ is the concentration of defect $D$ in charge state $q$.

To determine the overall non-radiative recombination kinetics for $V_{\mathrm{Cd}}$ in intrinsic CdTe, we first calculate the self-consistent Fermi level and concentrations of all native defects (using the calculated formation energies). For this, we assume a typical anneal temperature of $600\,^{\circ}\mathrm{C}$,[S68] in a Te-rich atmosphere with no impurity species present, producing a vacancy concentration $\left[V_{\mathrm{Cd}}\right] = 1.3 \times 10^{13}\,\mathrm{cm}^{-3}$. Upon quenching to room temperature ($\mathrm{T} = 300\,\mathrm{K}$) with all intrinsic defect concentrations 'frozen-in', we obtain a $p$-type Fermi level $0.14\,\mathrm{eV}$ above the VBM, with the majority of $V_{\mathrm{Cd}}$ in the $-1$ ($> 95\%$) and $-2$ ($\sim 4\%$) charge states — a consequence of total-energy minimization under the constraint of charge neutrality.

The overall non-radiative recombination kinetics upon photo-illumination are extremely rapid, following the $V_{\mathrm{Cd}}^{-1} \rightarrow V_{\mathrm{Cd}}^0{}_{\mathrm{Bipolaron}} \rightarrow V_{\mathrm{Cd}}^0{}_{\mathrm{Te\,Dimer}} \rightarrow V_{\mathrm{Cd}}^{-1}$ cycle in Figure S12. Due to the large hole concentrations in the $p$-type material ($p = 1.02 \times 10^{-17}\,\mathrm{cm}^{-3}$), the vast majority of Cd vacancies end up in the $V_{\mathrm{Cd}}^0{}_{\mathrm{Te\,Dimer}}$ state under photo-illumination, with electron capture by this defect species representing the rate-limiting step:

$$R_{Total} \;\simeq\; R_n^0 \;=\; C_n^0 \left[V_{\mathrm{Cd}}^0{}_{\mathrm{Te\,Dimer}}\right] n$$

With this recombination behavior, the minority carrier lifetime is limited to $29\,\mathrm{ns}$ and the maximum achievable photovoltaic efficiency is reduced from a Shockley-Quiesser limit of $32.1\,\%$ to a 'trap-limited conversion efficiency' (TLC)[S66] of $26.7\,\%$ (based on the bulk electronic properties, excluding interfacial effects and assuming perfect step-function absorption). This massive drop in efficiency demonstrates that the cadmium vacancy can act as a 'killer' defect center in intrinsic $p$-type CdTe. Moreover, this is a clear testament to the importance of chlorine treatment, strategic impurity doping and Cd-rich growth environments in the fabrication of high efficiency CdTe devices,[S68–S81] all of which contribute to either the

S-24

45

reduction or passivation of cadmium vacancies.

## S8.6 Comparison with Experimental and Theoretical Literature

The 1D configuration coordinate model is an approximation to multidimensional energy surfaces and combinations of normal-mode lattice vibrations.[S59,S60] Consequently, the calculated energy barriers and capture coefficients should be considered as upper and lower bounds of the true values, respectively. Regardless, the extremely small capture cross-sections of $\sigma < 10^{-20}\,cm^2$ for $\sigma_n^{-1}$, $\sigma_{p,\,Te\,Dimer}^{-1}$ and $\sigma_{n,\,Bipolaron}^0$ almost certainly rule out their experimental detection.

Kremer and Leigh[S37] and Scholz et al.[S38] reported hole traps in $p$-type CdTe, possibly associated with cadmium vacancies, at $0.32\,eV$ and $0.29\,eV$ above the VBM, with capture cross-sections of $\sigma_p(T = 250K) = 1.3 \times 10^{-18}\,cm^2$ and $\sigma_p(T = 215K) = 6 \times 10^{-18}\,cm^2$ respectively, using Deep Level Transient Spectroscopy (DLTS). We tentatively propose the negative-U (2-/0) $V_{Cd}$ level as the atomic origin in these cases, due to the close alignment with our calculated $V_{Cd}^{-2}$ hole capture cross-section $\sigma_p^{-2}(215 < T < 250K) = 2.3 \times 10^{-18}\,cm^2$ (the rate-limiting step in the $V_{Cd}^{-2} \rightarrow V_{Cd}^0$ transition) and thermal ionization energy ($0.35\,eV$). A defect complex involving $V_{Cd}^{-2}$ is another possible origin of this level.

Several groups have also reported a deep trap associated with the cadmium vacancy in the range $0.43$-$0.54\,eV$ above the VBM, using Photo-Induced Current Transient Spectroscopy (PICTS)[S82] and DLTS measurements,[S36,S38,S80,S82–S84] with capture cross-sections in the range $\sigma_p(T \sim 250K) = 1.6 \times 10^{-14} - 1.1 \times 10^{-13}\,cm^2$. These experimental values align with our calculated thermodynamic transition level $(-/0)_{Te\,Dimer}$ ($\Delta E = 0.483\,eV$) and capture cross-section ($\sigma_p(T = 188K) = 1.9 \times 10^{-13}\,cm^2$, for the $V_{Cd}^{-1} \rightarrow V_{Cd}^0{}_{Bipolaron}$ transition — the rate-limiting step for $V_{Cd}^{-1} \rightarrow V_{Cd}^0{}_{Te\,Dimer}$) for the $(-/0)$ $V_{Cd}$ level, and so we suggest this defect center as the origin of this deep trap.

Both Szeles et al.[S69] and Rakhshani and Makdisi[S85] also reported a $V_{Cd}$-associated hole trap at $0.42$-$0.46\,eV$ and $0.49\,eV$ above the VBM respectively, though their measured cap-

ture cross-sections of $\sigma_p(T = 188K) = 2 \times 10^{-16}\,\text{cm}^2$ and $\sigma_p(T = 380K) = 3.3 \times 10^{-17}\,\text{cm}^2$ are quite different to other experimentally reported values and our calculated values. The $\text{Te}_{\text{Cd}}$ antisite defect has been found both in this and other research,[S44] to be a low-energy defect in Te-rich CdTe and to exhibit DX-center type relaxation in certain charge states, such that $\text{Te}_{\text{Cd}}{}^{-2} \simeq (\text{Te-Te})_i{}^{-2}$ (split interstitial) $+ V_{\text{Cd}}$. Thus we propose a deep DX-center $\text{Te}_{\text{Cd}}$ level as a likely origin of this apparent discrepancy, which will be further analyzed in following work.

Using an admittance spectroscopy technique, Reislöhner et al.[S86] found a thermal ionization energy of $0.23(3)\,\text{eV}$ for a hole trap in CdTe attributed to the $(-/2-)$ $V_{\text{Cd}}$ level, in good correspondence with our predicted value of $\Delta E = 0.22\,\text{eV}$ (Figure S9a).

Yang et al.[S87] also calculated carrier capture coefficients for three $V_{\text{Cd}}$ traps. However, these values unfortunately are not comparable to our results, as they did not find the equilibrium $C_{3v}$ configuration for $V_{\text{Cd}}{}^{-1}$, which drastically affects the carrier capture PESs and thus overall recombination behavior, and employed the harmonic approximation in the charge capture model. Likewise, Krasikov et al.[S88] calculated capture cross-sections for $V_{\text{Cd}}$ in CdTe, which again are several orders of magnitude off our predicted behavior, due to the exclusion of the Te dimer and hole polaron configurations for $V_{\text{Cd}}{}^0$ and $V_{\text{Cd}}{}^{-1}$ and use of the harmonic approximation.

## S8.7 Considerations for Accurate Calculations of Recombination Kinetics

In addition to demonstrating the impact of $V_{\text{Cd}}$ on CdTe PV performance, these results yield important considerations for the accurate modeling of defect-mediated recombination in photovoltaic materials.

Firstly, the correct location of equilibrium defect structures (such as the $C_{3V}$ hole polaron $V_{\text{Cd}}{}^{-1}$ and Te dimer $V_{\text{Cd}}{}^0$ states) is essential, not just for accurate prediction of charge transition energy levels and negative-U behavior, but also for carrier capture kinetics.

Secondly, by performing a relatively dense sampling of the defect PES across the configuration landscape (Figure 6), we have incorporated the effects of anharmonicity and ensured the accuracy of the fitted potentials. Often this is avoided by the use of the harmonic approximation, due to the associated increase in computational cost. While appropriate for the $(2 - / -)$ and $(- / 0)_{\text{Bipolaron}}$ centers, the approximation of harmonic PESs grossly fails for the $(- / 0)_{\text{Te Dimer}}$ level — a direct result of strong local distortion and anharmonicity due to Te dimer formation. However, our results demonstrate the necessity of this procedure for defects which undergo strong lattice relaxation upon charge transition (such as the Te dimer $V_{\text{Cd}}{}^0$ arrangement), as demonstrated in other recent works.[S60,S89] To illustrate the critical dependence of predicted non-radiative recombination activity on these considerations, we note that, without both of them, negligible electron capture would be expected at $V_{\text{Cd}}$ ($C_n(T = 300K) < 10^{-20}\,\text{cm}^3/\text{s}$ for the $(2 - / -)$ and $(- / 0)_{\text{Bipolaron}}$ levels — Table S3) and thus negligible electron-hole recombination.

Finally, it is important to note that the use of solely the *equilibrium* structures for each defect charge state would lead to completely erroneous predicted carrier capture kinetics in this case. In fact, the overall non-radiative recombination rate at $V_{\text{Cd}}$ would be reduced by approximately seven orders of magnitude (Table S3), compared to the inclusion of metastable (*i.e.* neutral bipolaron) configurations, leading to the spurious prediction of negligible impact of $V_{\text{Cd}}$ on CdTe device efficiencies. The requirement of excited states in the calculation of carrier capture behavior, in order to yield results matching experimental observations, has been noted in the literature in recent years.[S56,S90]

Overall, we have provided a powerful demonstration of the necessity to obtain correct equilibrium defect structures and include the effects of both metastable configurations and anharmonic energy surfaces for the accurate calculation of non-radiative recombination rates in photovoltaic materials.

## S9    Experimental Identification of Tellurium dimerization

Experimental verification of tellurium dimerization at point defects in CdTe (which we also predict for other low-energy CdTe defects; tellurium-on-cadmium antisites $Te_{Cd}$ and tellurium interstitials $Te_i$) would provide valuable further evidence of the previously-hidden impact of this species on electron-hole recombination in CdTe and the accuracy of state-of-the-art defect modelling techniques. While Te dimers have been experimentally observed at grain boundaries and surfaces in CdTe,[S29,S30] to the knowledge of the authors, there have been no experimental reports of this species at point defects thus far. We believe this is a consequence of the recency of their prediction, the inherent difficulty in structural characterization of low concentration, diamagnetic, intrinsic defects and the fact that CdTe film growth is typically performed using methods which minimise defect formation (and thus non-radiative recombination). We suggest that nuclear magnetic resonance (NMR) — 8 % naturally-abundance of NMR-active Te — or highly-sensitive IR/Raman vibrational spectroscopy on an ion-irradiated or vacuum-annealed sample of CdTe under Te-rich conditions (to induce sufficiently large defect concentrations) could possibly allow for the identification of this structural motif.

Two other options include the use of (1) Mössbauer Spectroscopy — from which the isomer shift and (possibly) nuclear splittings could evidence the Te - Te dimer state — or possibly (2) Optically Detected Magnetic Resonance (ODMR) — to measure the $V_{Cd}{}^0$ (diamagnetic) $\rightarrow V_{Cd}{}^{-1}$ (paramagnetic) optical transition energy, vibrational relaxation and luminescence behavior.

## S10 Additional Electronic Densities of States

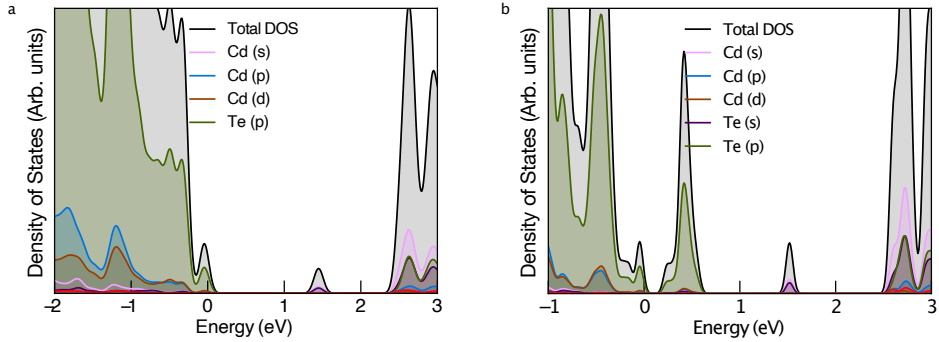

Figure S13: Electronic density of states for a CdTe supercell containing (**a**) $V_{Cd}^{-2}$ and (**b**) the metastable $C_{2v}$ bi-polaron solution for $V_{Cd}^{0}$ (showing the unoccupied Te $p$ hole states just above the VBM. Fermi level set to 0 eV.

## S11 Chemical Potentials

Being a binary semiconductor, CdTe has a simple phase stability diagram, bounded by pure metallic Cd ($\mu_{Cd} = 0$) and Te metalloid ($\mu_{Te} = 0$).

Table S4: Calculated chemical potential (free energy of formation) of CdTe, compared to experiment and other theoretical values in literature.

| Functional | $\mu_{\mathbf{CdTe}}$ |
|---|---|
| ***Experiment***[S91] | $-1.17\,\text{eV}$ |
| **HSE**(34.5 %)(**This Work**) | $-1.25\,\text{eV}$ |
| HSE (33 %) & FERE[S23] | $-1.26\,\text{eV}$ |
| PBE0[S92] | $-1.13\,\text{eV}$ |
| GGA+U$_{SIC}$[S92] | $-1.78\,\text{eV}$ |



\end{document}